\documentclass{aa}
\usepackage{subfig}        
\usepackage[T1]{fontenc}
\usepackage{fixltx2e,longtable,ifpdf,amsmath,amssymb,natbib,textcomp}
\ifpdf
  \usepackage{aeguill}
  \usepackage[pdftex]{graphicx,color}
\else
  \usepackage[dvips]{graphicx,color}
\fi

\usepackage{multirow,bigdelim,bigstrut,rotating}
\newcommand\T{\rule{0pt}{2.6ex}}       
\newcommand\B{\rule[-1.2ex]{0pt}{0pt}} 

\newcommand{\teff}{\ensuremath{T_\mathrm{eff}}}
\newcommand{\logg}{\ensuremath{\log{g}}}

\newcommand{\msun}{\ensuremath{\mbox{M}_{\odot}}}
\newcommand{\rsun}{\ensuremath{\mbox{R}_{\odot}}}
\newcommand{\lsun}{\ensuremath{\mbox{L}_{\odot}}}

\newcommand{\vsini}{\ensuremath{v\sin{i}}}
\newcommand{\vesini}{\ensuremath{v_\mathrm{e}\sin{i}}}

\newcommand{\vcpsini}{\ensuremath{v_1{\rm{(c)}}\sin{i}}}
\newcommand{\vcssini}{\ensuremath{v_2{\rm{(c)}}\sin{i}}}
\newcommand{\vcxsini}{\ensuremath{v_{\rm i}{\rm{(c)}}\sin{i}}}

\newcommand{\hei} {\mbox{He$\;$\sc{i}}}
\newcommand{\heii} {\mbox{He$\;$\sc{ii}}}

\newcommand{\siiii} {\mbox{Si$\;$\sc{iii}}}

\newcommand{\ciii} {\mbox{C$\;$\sc{iii}}}

\newcommand{\nii} {\mbox{N$\;$\sc{ii}}}
\newcommand{\niii} {\mbox{N$\;$\sc{iii}}}

\newcommand{\fstwo} {\ensuremath{\mathcal{F}_2}}
\newcommand{\fsone} {\ensuremath{\mathcal{F}_1}}

\newcommand{\kms} {\ensuremath{\mbox{km}\;\mbox{s}^{-1}}}

\newcommand{\pa}  {\ensuremath{\phantom{1}}}
\newcommand{\pb}  {\ensuremath{\phantom{]}}}
\newcommand{\pc}  {\ensuremath{\phantom{,}}}
\newcommand{\ps}  {\ensuremath{\phantom{*}}}
\newcommand{\px}  {\ensuremath{\phantom{\le}}}

\def\5{\footnotesize V\normalsize}
\def\4{\footnotesize IV\normalsize}
\def\3{\footnotesize III\normalsize}
\def\2{\footnotesize II\normalsize}
\def\1{\footnotesize I\normalsize}

\begin{document}
\title{The VLT-FLAMES Tarantula Survey}
\subtitle{XXIII: two massive
  double-lined  binaries in 30~Doradus\thanks{Based on observations obtained at the
    European Southern Observatory Very Large Telescope (VLT) as part of programmes
    \mbox{182.D-0222} and 090.D-0323}}

\author{Ian~D.~Howarth,\inst{1} 
P.~L.~Dufton,\inst{2} P.~R.~Dunstall,\inst{2}
C.~J.~Evans,\inst{3}
L.~A.~Almeida,\inst{4,5}
A.~Z.~Bonanos,\inst{6}
J.~S.~ Clark,\inst{7}
N.~Langer,\inst{8} 
H.~Sana,\inst{9} 
S.~Sim\'on-D\'iaz,\inst{10,11}
I.~Soszy\'nski,\inst{12} and
W.~D.~Taylor\inst{3}} 
\institute{ Department of Physics \& Astronomy, University College London, Gower Street, London, WC1E 6BT, UK
\and Astrophysics Research Centre, Queen's University Belfast, BT7 1NN, Northern Ireland, UK
\and UK Astronomy Technology Centre, Royal Observatory Edinburgh, Blackford Hill, Edinburgh, EH9 3HJ, UK
\and Department of Physics \& Astronomy, Johns Hopkins University, 
Bloomberg Center for Physics and Astronomy, Room 520, 
3400 N.\ Charles Street, Baltimore, MD 21218, USA
\and Instituto de Astronomia, Geof\'isica e Ci\^encias, Rua do
Mat\~ao 1226, Cidade Universit\'aria S\~ao Paulo, SP, Brasil, 05508-090
\and IAASARS, National Observatory of Athens, GR-15236, Penteli, Greece
\and Department of Physics \& Astronomy, The Open University, Walton Hall, Milton Keynes, MK7 6AA, UK
\and Argelander Institut f\"ur Astronomie der Universit\"at Bonn, Auf dem H\"ugel 71, 53121, Bonn, Germany
\and ESA/STScI, 3700 San Martin Drive, Baltimore, MD 21218, USA
\and Instituto de Astrof\'isica de Canarias, E-38200 La Laguna, Tenerife, Spain
\and Departamento de Astrof\'isica, Universidad de La Laguna, E-38205 La Laguna, Tenerife, Spain
\and Warsaw University Observatory, Al. Ujazdowskie 4, 00-478 Warszawa, Poland
%
}
       
\offprints{Ian Howarth\\ \email{i.howarth@ucl.ac.uk }}

\date{Accepted for publication in A\&A, 2015/8/24}

\abstract{} {We investigate the characteristics of two newly
  discovered short-period, double-lined, massive binary systems,
  VFTS~450 \mbox{(O9.7$\;$II--Ib\,+\,O7::)} and VFTS~652
  \mbox{(B1$\;$Ib{\,+\,}O9:$\;$III:)}.}  {We perform model-atmosphere
  analyses to characterise the photo\-spheric properties of both
  members of each binary (denoting the `primary' as the
  spectroscopically more conspicuous component).  Radial velocities
  and optical photo\-metry are used to estimate the binary-system
  parameters.}  {We estimate $\teff=27$~kK, $\logg=2.9$ (cgs) for the
  VFTS~450 primary spectrum (34~kK, 3.6:\ for the secondary spectrum); and $\teff =
  22$~kK, $\logg=2.8$ for the VFTS~652 primary spectrum (35~kK, 3.7: for the
  secondary spectrum).  Both primaries show surface nitrogen enrichments (of
  more than 1~dex for VFTS~652), and probable moderate oxygen
  depletions relative to reference LMC abundances.  We determine
  orbital periods of 6.89~d and 8.59~d for VFTS~450 and VFTS~652,
  respectively, and argue that the primaries must be close to filling
  their Roche lobes.  Supposing this to be the case, we estimate
  component masses in the range $\sim$20--50\msun.}  {The secondary
  spectra are associated with the more massive components, suggesting
  that both systems are high-mass analogues of classical Algol systems, undergoing
  case-A mass transfer.  Difficulties in reconciling the spectroscopic
  analyses with the light-curves and with evolutionary considerations
  suggest that the secondary spectra are contaminated by (or arise in)
  accretion disks.}


\authorrunning{I.D.~Howarth et al.}
\titlerunning{Two massive binaries}

\maketitle

\section{Introduction}
\label{s_intro}

Massive, luminous stars are of interest for the role that they play in
galactic chemical evolution; the environmental impact they have
through mechanical and radiative energy input to their surroundings;
and as tracers of recent star formation.  However, while there have been considerable
advances in modelling their spectra, \emph{direct} determinations of their
fundamental parameters are relatively few, because of the
scarcity of suitable double-lined eclipsing binary systems (cf., e.g.,
\citealt{bon09a} and references therein).


Multi-epoch spectroscopy from the VLT-FLAMES Tarantula Survey (VFTS;
\citealt{eva11}) of the OB-star population of 30 Doradus, in the Large
Magellanic Cloud (LMC), has led to the discovery of a number of
systems showing radial-velocity variations that appear to be
consistent with binary motion (\citealt{sana13, dunstall15}).  These
systems offer an important opportunity to better understand
the physical properties of stars in the upper Hertzsprung--Russell
diagram, as exemplified by the VFTS study of R139 by \citeauthor{tay11}
(\citeyear{tay11}).

\begin{figure}
\begin{center}
\includegraphics[scale=0.30]{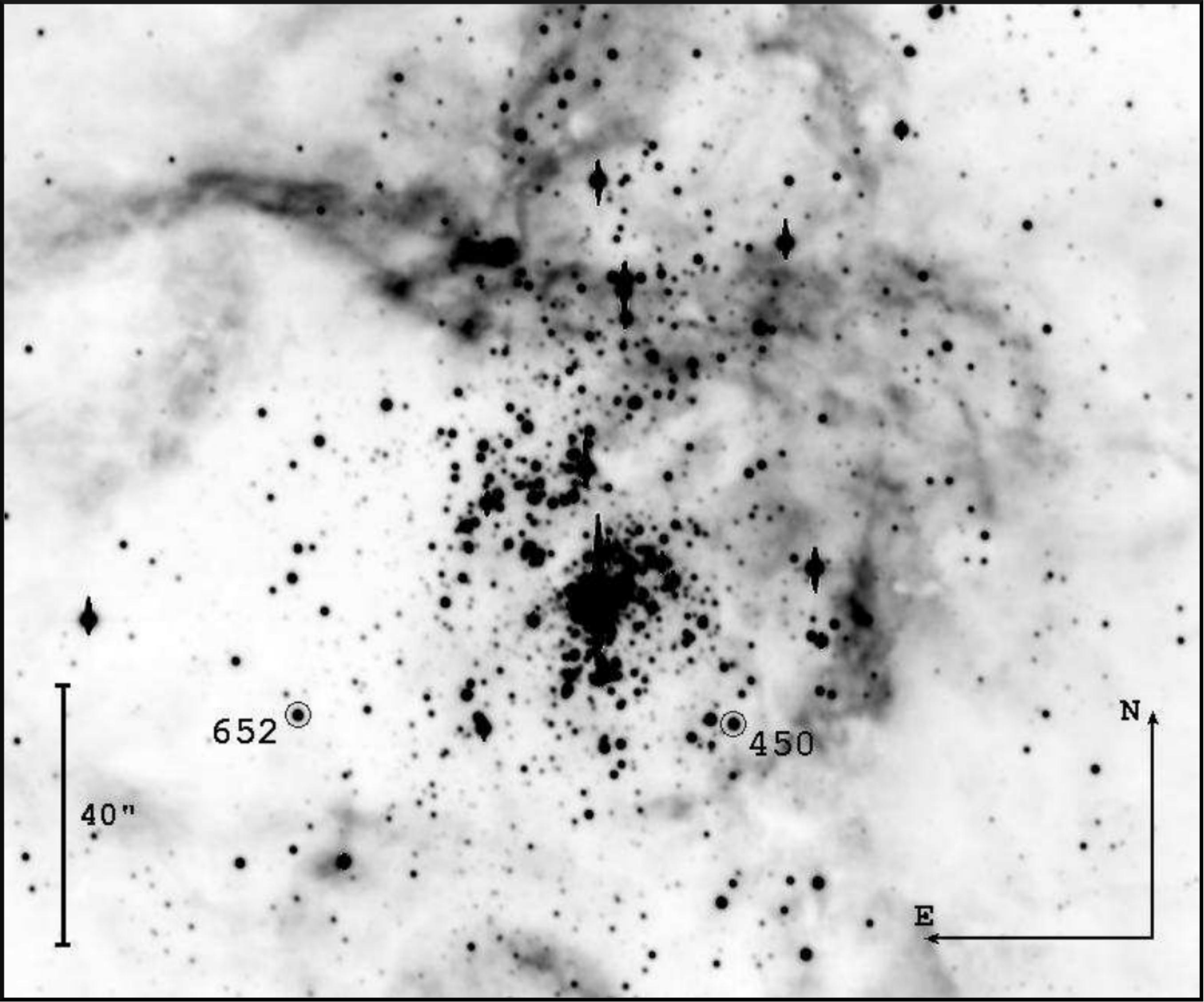}
\caption{The central region of 30~Dor, showing R136 and the locations
  of VFTS~450 and VFTS~652.}
\label{f_locations}
\end{center}
\end{figure}

Here we discuss two newly identified double-lined radial-velocity
variables discovered in the VFTS: nos.~450 and 652 (stars~50 and~5 in 
\citealt{mel85}).
The primary velocity amplitudes are among the largest measured in the
VFTS dataset, and the number of spectroscopic observations available
makes it possible to undertake full orbital analyses (rather than mere
detections of variability). Each system also shows orbital photometric
variability. Both are relatively close to the core of 30~Dor, with
radial distances of 0{\farcm}47 and 0{\farcm}83 from R136
(Fig.~\ref{f_locations}), corresponding to projected distances of 6.8
and 12.0\,pc, respectively.

The paper is organized as follows: Section~\ref{s_obs} summarizes the
data, and  the binary characteristics are
\mbox{examined} in Section~\ref{s_analysis}.
A model-atmosphere analysis is described in
Section~\ref{s_stellar}, and simple models of the
systems are constructed in Section~\ref{sec:sys}.  Throughout the
paper we adopt the convention that the `primary' in each system is the
star with the stronger optical absorption-line spectrum (though we
shall argue that this is probably not the more
massive component).

\goodbreak

%

\begin{table}[htpb]
  \caption{Characteristics of spectroscopic observations.}
\begin{center}
\begin{tabular}{lccc}
\hline\hline
\multicolumn{1}{c}{Wavelength}&\rule{0pt}{8pt}$\lambda$ range&\rule{0pt}{9pt}
Resolving & Typical \\
\multicolumn{1}{c}{setting} & (\AA) & power $R$ & S:N \\
\hline
\rule{0pt}{9pt}LR02 & 3960--4560 & {\pa}7\,000 & 100--300	\\
LR03 & 4505--5050 & {\pa}8\,500	& 150  \\
UVES 520 & {\pc}4175--5155, & 53\,000 &  20--40\\
         & 5240--6200 \\
\hline
\end{tabular}
\end{center}
\label{t_obs}
\end{table}

\addtocounter{table}{2}
\begin{table*}
\caption{Basic observed properties.}
\begin{center}
\begin{tabular}{clccccccccccc}
\hline\hline
\rule{0pt}{9pt}VFTS & \multicolumn{1}{c}{Spectral Type} & $V$ & $B -
V$  & $J$ & $H$ & $K_{\rm s}$ & [3.6] & [4.5] & [5.8] &$<I_{\rm
  C}>$&$<V-I_{\rm C}>$\\
\hline
\rule{0pt}{9pt}450  
    & O9.7$\;$II--Ib{\,+\,}O7:: & 13.60 & 0.20 & 13.08 & 12.91 & 12.89 & 11.00 & 10.65 & 10.52 &13.26&+0.19\\
652 & B1$\;$Ib{\,+\,}O9:$\;$III: & 13.88 & 0.20 & 13.40 & 13.28 & 13.22 & 12.97 & $\cdots$ & 12.73&13.63&+0.36  \\
\hline
\end{tabular}
\tablefoot{Photometry follows \citet{bon09b}.  The primary source for
  the $B$, $V$ photometry is \citet{sel99}; their observations were
  obtained over only $\sim$10~minutes, so the $(B-V)$ colours are
  insensitive to orbital variability.
%
%
%
\emph{JHK} results are from IRSF (the InfraRed Survey Facility;
\citealt{kat07}), and the mid-IR photo\-metry from \textit{Spitzer}
`SAGE' Legacy Science Program \citep{mei06}.  Measurement
uncertainties are $\lesssim$0{\fm}05, excepting the [5.8] magnitude
for VFTS~450 ($\pm$0{\fm}15), but both stars are variable with
amplitudes of $\sim$0{\fm}2 (Fig.~\ref{fig:photom}).  The last two
columns are average results from our OGLE photometry
($\S$\ref{s_ogle}).}
\end{center}
\label{t_photo}
\end{table*}

\begin{figure*}
\begin{center}
\includegraphics[angle=270, scale=0.71]{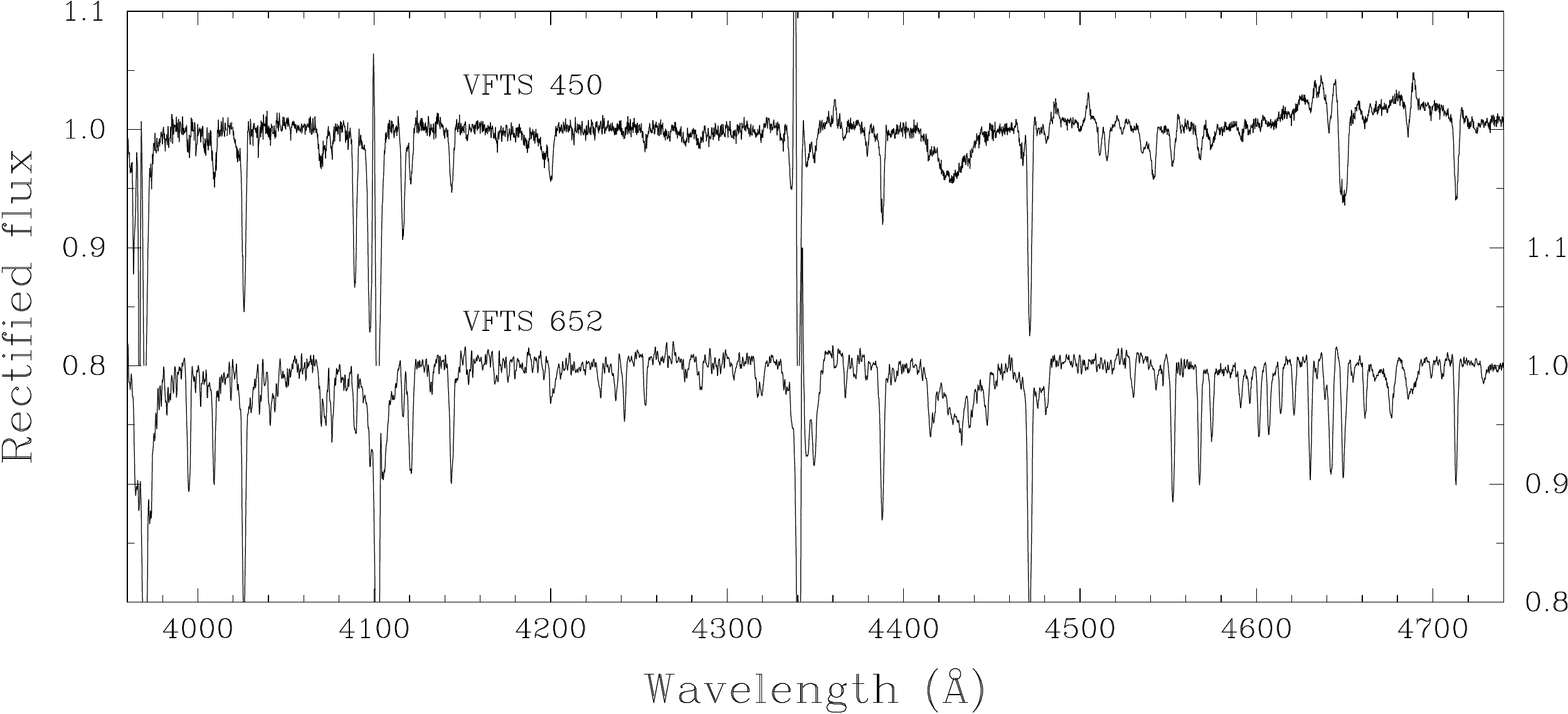} 
\caption{Rectified blue-region spectra of VFTS~450 and 652, velocity
  shifted to the rest frame of the primary.  The data are LR02 and
  LR03 spectra from MJD 54748, 54810 (VFTS~450; $\phi \simeq 0.12$
  from the circular-orbit ephemeris in Table~\ref{tab:specorb}) and
  56294, 54808 (VFTS~652, $\phi \simeq 0.66$), merged at
  $\sim$4560\AA.  Secondary spectra are offset by ca.\ +260,
  $-$150~\kms\ (VFTS~450, 652, respectively).  
Narrow Balmer emission
  is nebular.}
\label{f_bluespec}
\end{center}
\end{figure*}

\begin{figure}
\begin{center}
\includegraphics[angle=270, scale=0.37]{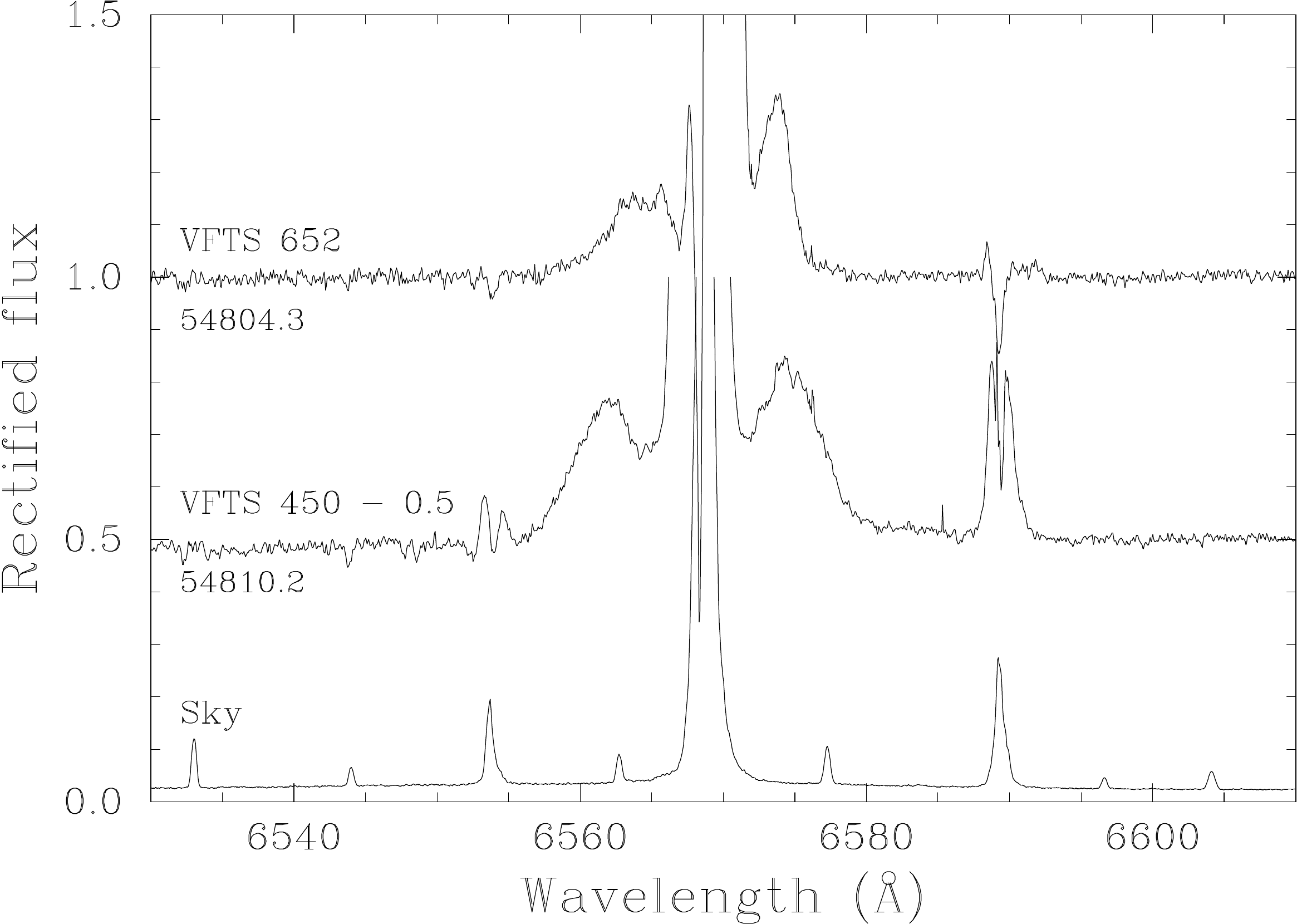} 
\caption{H$\alpha$ spectra, labelled with MJDs of mid-observation.
  The subtracted sky spectrum is shown for reference, and illustrates
  the nebular contamination, which varies on small spatial scales;
  correction for this nebular emission in the stellar spectra is
  generally poor, in particular in the residual core
  H$\alpha$ emission, although the extended double-peaked emission is
  real.}
\label{f_redspec}
\end{center}
\end{figure}

\begin{figure}
\begin{center}
\includegraphics[angle=270, scale=0.55]{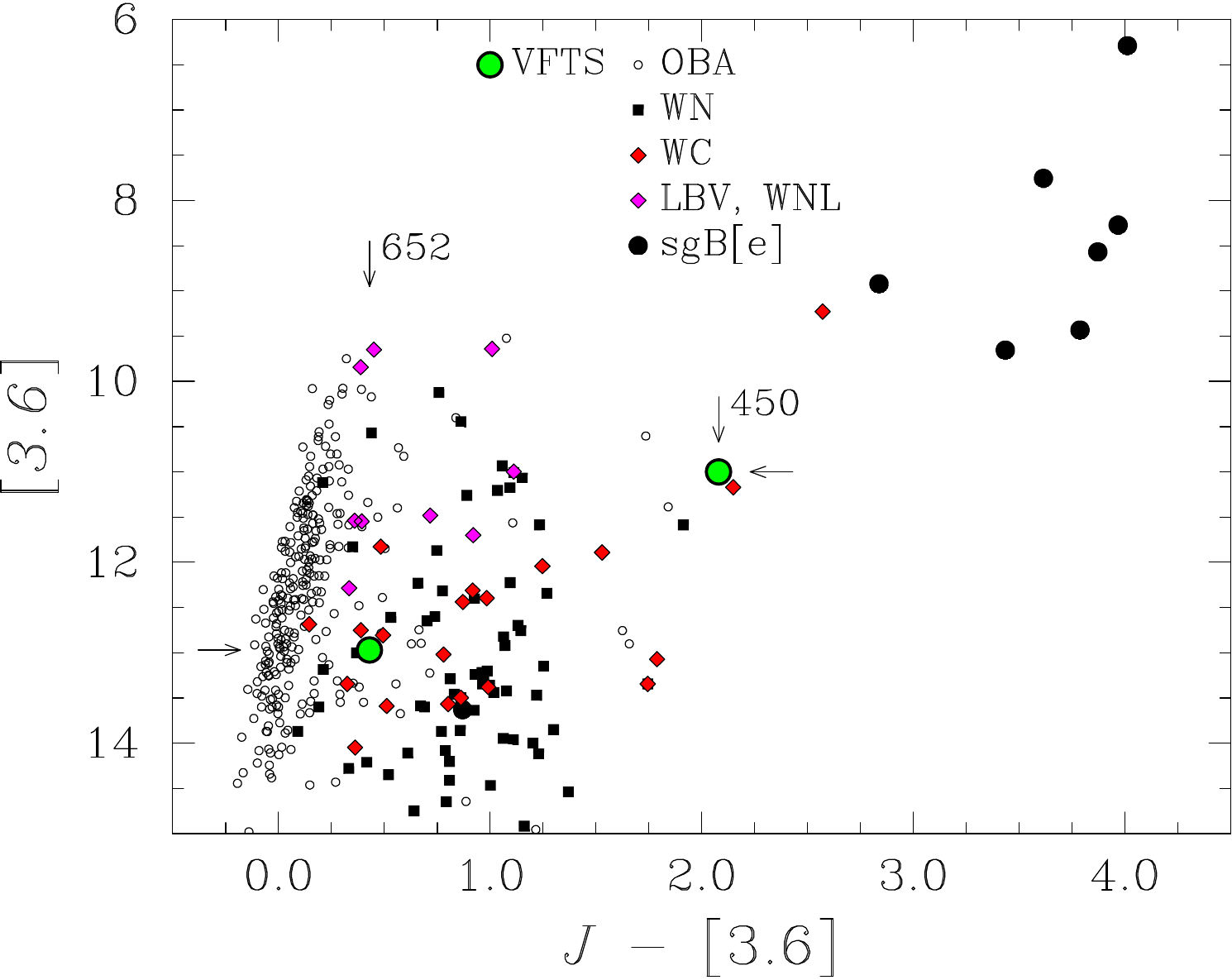} 
\caption{Colour--magnitude diagram, comparing VFTS~450 and
  652 to selected LMC emission-line stars
  and OBA super\-giants  (after \citealt{bon09b}).}
\label{f_near_IR}
\end{center}
\end{figure}

\section{Observations}
\label{s_obs}
\subsection{Optical Spectroscopy}
\label{s_obs_spec}

Initial spectroscopic data were obtained as part of the VFTS
\citep{eva11}, using the Fibre Large Array Multi-Element Spectrograph
(FLAMES; \citealt{pas02}) on the Very Large Telescope, primarily with
the Giraffe spectrograph, but with supplementary data from the
Ultraviolet and Visual Echelle Spectrograph (UVES).  These observations were
obtained in the 2008/9 and 2009/10 observing seasons; additional
Giraffe spectroscopy was secured as part of a binary-monitoring campaign
of VFTS targets between 2012 Oct and 2013 Mar.  

Table~\ref{t_obs} summarizes the basic instrumental characteristics; a
full account of the observations and data reduction is given by
\citet{eva11}.  Logs of the individual blue-region spectra, which are
the principal focus of this paper, are given in
Tables~\ref{t_full_obs1} and~\ref{t_full_obs2} (on-line).
Representative spectra are shown in Fig.~\ref{f_bluespec}.

\subsubsection{Spectral types}

Spectral types previously determined from the VFTS spectra are
O9.7$\;$III:{\,+\,}O7:: and B2$\;$Ip{\,+\,}O9$\;$III:
(VFTS~450, VFTS~652; \citealt{wal14}).  \citet{mel85} gives O9.5$\;$I
and O9.5$\;$V~pec (``binary?'') for VFTS~450 and 652, respectively,
while \citet{wal97} report ON9:$\;$I and B2$\;$Ib.

Our review of the more extensive dataset discussed here, including
examination of the disentangled component spectra presented in
Section~\ref{sec:disentangle}, broadly supports the \citet{wal14}
classifications, but the clear presence of \ion{Si}{iv} $\lambda$4089
and $\lambda$4116 in the primary spectrum of VFTS~652 leads us to
revise its classification to B1$\;$Ib.  The crucial
\ion{He}{i}~$\lambda{4471}$ classification line suffers significant nebular
contamination in the disentangled secondary spectra, admitting the
possibility of an O8 (or, conceivably, O7) secondary spectrum for this
target.

Our rectification of the VFTS~450 spectra leaves a broad, shallow
emission feature spanning $\lambda\lambda$4640, 4686 (\ciii/\niii,
\heii; Fig.~\ref{f_bluespec}).  We have investigated, and rejected,
possible instrumental origins, including contamination by the nearby
WR star Brey~79 (3\farcs5 distant).  While such features are not
widely reported, and are easily overlooked, they are not unprecedented
in late-O supergiants (e.g., $\alpha$~Cam, O9$\;$Ia;
\citealt{Wilson58}); 
this suggests the possibility of a brighter luminosity class for the
VFTS~450 primary than previously inferred from VFTS data.  The
intensity of the \ion{Si}{iv} lines compared to \ion{He}{i}
$\lambda$4026 also indicates a somewhat more luminous type
(cf.~Table 6 of \citealt{sot11}).  The arbitrary intensity
scaling of the disentangled spectrum hampers a precise assignment, but
we revise the previous
classification for the VFTS~450 primary spectrum to
O9.7$\;$II--Ib.\footnote{We recall the convention that `II--Ib' is
  to be read as indicating a range of uncertainty, whereas `Ib--II'
  would indicate a precise interpolated luminosity class.}
Our adopted spectral types are incorporated into Table~\ref{t_photo}.

\subsubsection{H$\alpha$ spectra} 

We have H$\alpha$ observations, shown in Fig.~\ref{f_redspec}, at a
single epoch for each system.  These spectra suffer from strong nebular
contamination which is poorly corrected by standard sky subtraction,
but nevertheless each star clearly shows broad, double-peaked
intrinsic emission.  Although the single-epoch spectra may not be
representative of typical behaviour, this emission morphology is
characteristic of 
interacting binaries,
rather than typical OB-star stellar-wind P-Cygni profiles.  Peak-to-peak
separations are $\sim$560 and 420~\kms\ for VFTS~450 and~652,
respectively, with full widths at continuum level about twice those
values.

\begin{figure}
\centering
\includegraphics[scale=0.56,angle=0]{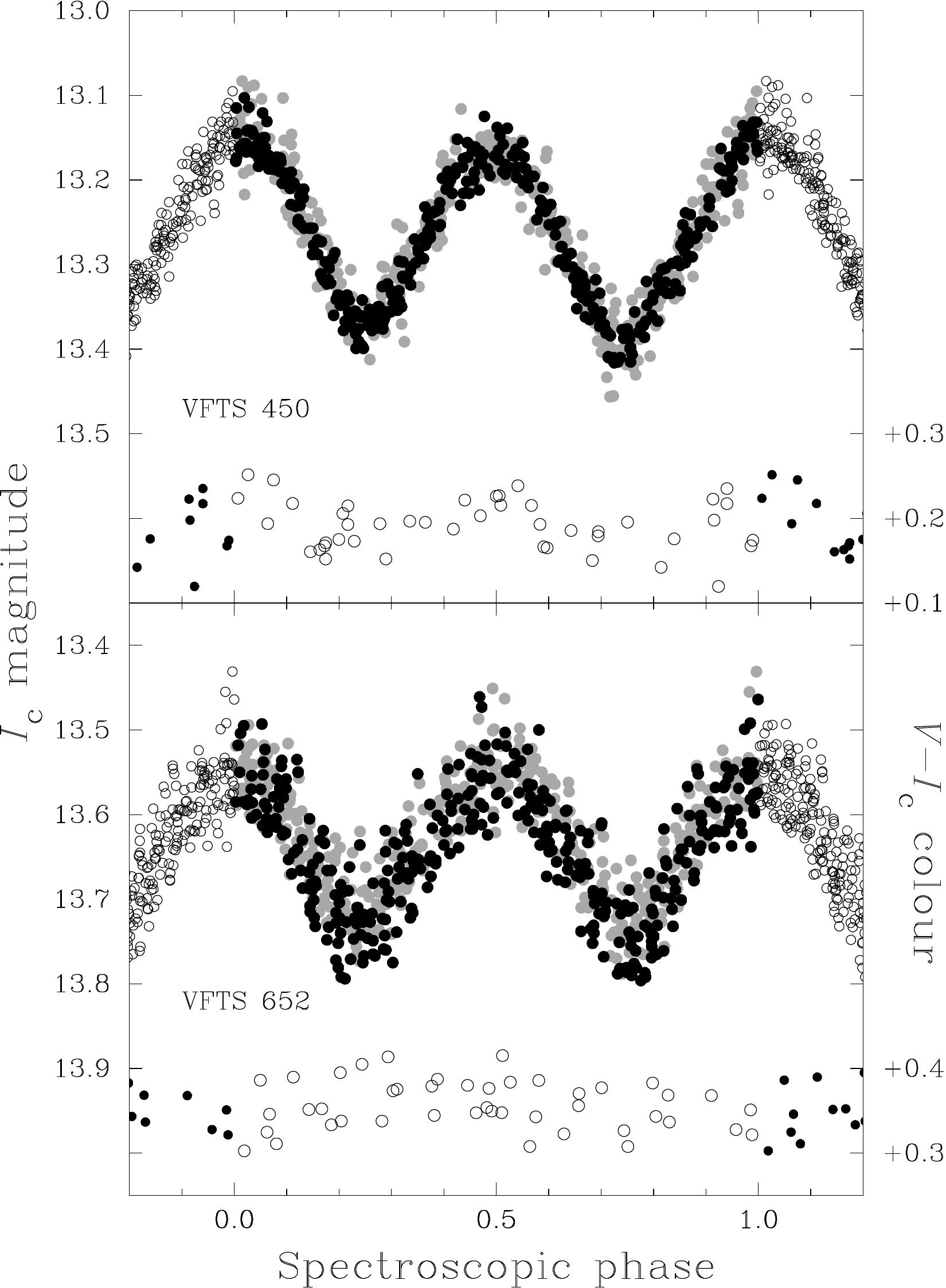}
\caption{OGLE photometry.  Phases are computed with respect to the
  photometric circular-orbit $T_0$ values and periods reported in
  Table~\ref{tab:specorb}.  OGLE-III and OGLE-IV magnitudes are shown
  in grey and black, respectively;  the OGLE-III results for VFTS~450
  have been offset by $-0{\fm}174$ (cf.~$\S$\ref{s_obs_photo2}).
The $V$- and $I_{\rm C}$-band measurements
  were not quasi-simultaneous, and OGLE-III colours have been computed
  from phase-binned $I_{\rm C}$ results, which are $\sim$10$\times$ as
  numerous as the $V$ measurements.}
\label{fig:photom}
\end{figure}

\subsection{Photometry}
\label{s_obs_photo2}

\subsubsection{OGLE photometry}
\label{s_ogle}

We have OGLE $I_{\rm C}$-band photometry from phases III and IV of the
project (cf.\ \citealt{uda08, uda15}), spanning 2001 October to 2009
April and 2010 March to 2014 March, respectively.  Each dataset for
each star consists of $\sim$400 observations.   We have also examined
the sparser OGLE-III $V$-band data.

VFTS~450 is located in a high-background region, and, as noted above, is only
$\sim$3{\farcs}5 arcsec from Brey~79 ($I_{\rm C} \simeq 12.7$).  The standard
OGLE-III Difference Image Analysis (DIA) pipeline is not optimal under
these circumstances, and we found that a profile-fitting extraction,
using \textsc{DoPhot} \citep{dophot}, resulted in reduced scatter.
Furthermore, the OGLE-IV $I_{\rm C}$ photometry for this target is
$\sim$0{\fm}17 brighter than the OGLE-III data (regardless of extraction
method). This is probably a consequence of the high background; while
it is difficult to be certain, we believe the OGLE-IV normalization to
be the more reliable.  Neither issue arises in the VFTS~652
results.

Both stars show orbital photometric variability, with full amplitudes
of $\sim$0{\fm}2. Periods were determined by using a date-compensated
discrete fourier transform (\citealt{ferraz81}), augmented with
least-squares fitting of a double sine wave. Results are included in
Table~\ref{tab:specorb}.   There are no significant differences in
periods determined from the OGLE-III, OGLE-IV, and combined datasets
(Table~\ref{tab:specorb} gives results from the combined $I_{\rm C}$-band data).
The phased OGLE photo\-metry is shown in
Fig.~\ref{fig:photom}.
The rms scatter about the mean curve for
VFTS~450, $\sim$0{\fm}023, is consistent with the probable measurement
uncertainties, but the larger scatter for VFTS~652, $\sim$0{\fm}036,
suggests significant intrinsic variability.

\subsubsection{IR photometry}
Representative visual--IR magnitudes for both stars, adapted from the
compilation by \citet{bon09b}, are listed in Table~\ref{t_photo}, and
are included in a $J-[3.6], [3.6]$ colour--magnitude \mbox{diagram} of
luminous LMC sources in Fig.~\ref{f_near_IR}.  VFTS~652 lies at the
red edge of the distribution of normal OBA super\-giants in this
figure (though this
displacement from the main grouping could possibly arise from
different phase sampling at $J$ and $[3.6]$ of the orbital
photo\-metric variability discussed in Section~\ref{sec:orb}).
However, at these wavelengths VFTS~450 has a substantial IR excess,
intermediate between those of typical Wolf-Rayet stars and super\-giant
B[e] stars.

\section{Spectroscopic orbits}
\label{s_analysis}

\subsection{Radial-velocity measurements}
\label{s_rvm}

\subsubsection{Primaries}

Radial
velocities for the primary components were reasonably straightforward to
measure using relatively unblended
\ion{Si}{iii}
and \ion{He}{i} 
lines.  We used the results of the model-atmosphere analyses reported
in Section~\ref{s_stellar} to identify suitable {\sc tlusty} models to
employ as templates in a cross-correlation analysis.  Results, which
are insensitive to the precise choice of model template, are
incorporated into Tables~\ref{t_full_obs1} and~\ref{t_full_obs2}
(on-line); the dispersion in velocities from different lines, and
residuals from the orbital solutions discussed in
Section~\ref{sec:orb}, are consistent with measurement errors of
$\lesssim$10~\kms.

\subsubsection{Secondaries}
\label{sec:sspec}

In order to measure the much weaker secondary spectra, we merged
LR02/LR03 spectra taken on any given night using a weighted mean, with
a sigma-clipping algorithm to exclude cosmic-ray events and other
flaws.  (Multiple observations taken at a given spectrograph setting
on any one night span $\lesssim$1{\%} of the orbital periods that we
report in Table~\ref{tab:specorb}, and may therefore be combined
without special procedures to compensate for binary motion.)
Uncertain corrections for echelle blaze render measurements in the
UVES spectra unreliable.  

The secondary velocities were measured by direct fitting of gaussians,
but the shallowness and breadth of the lines make the results quite
sensitive to the adopted rectification.  Repeat measurements and
residuals to model fits are both consistent with typical measurement
errors of $\sim$35~\kms.

\begin{figure}
\begin{center}
\includegraphics[angle=0, scale=0.55]{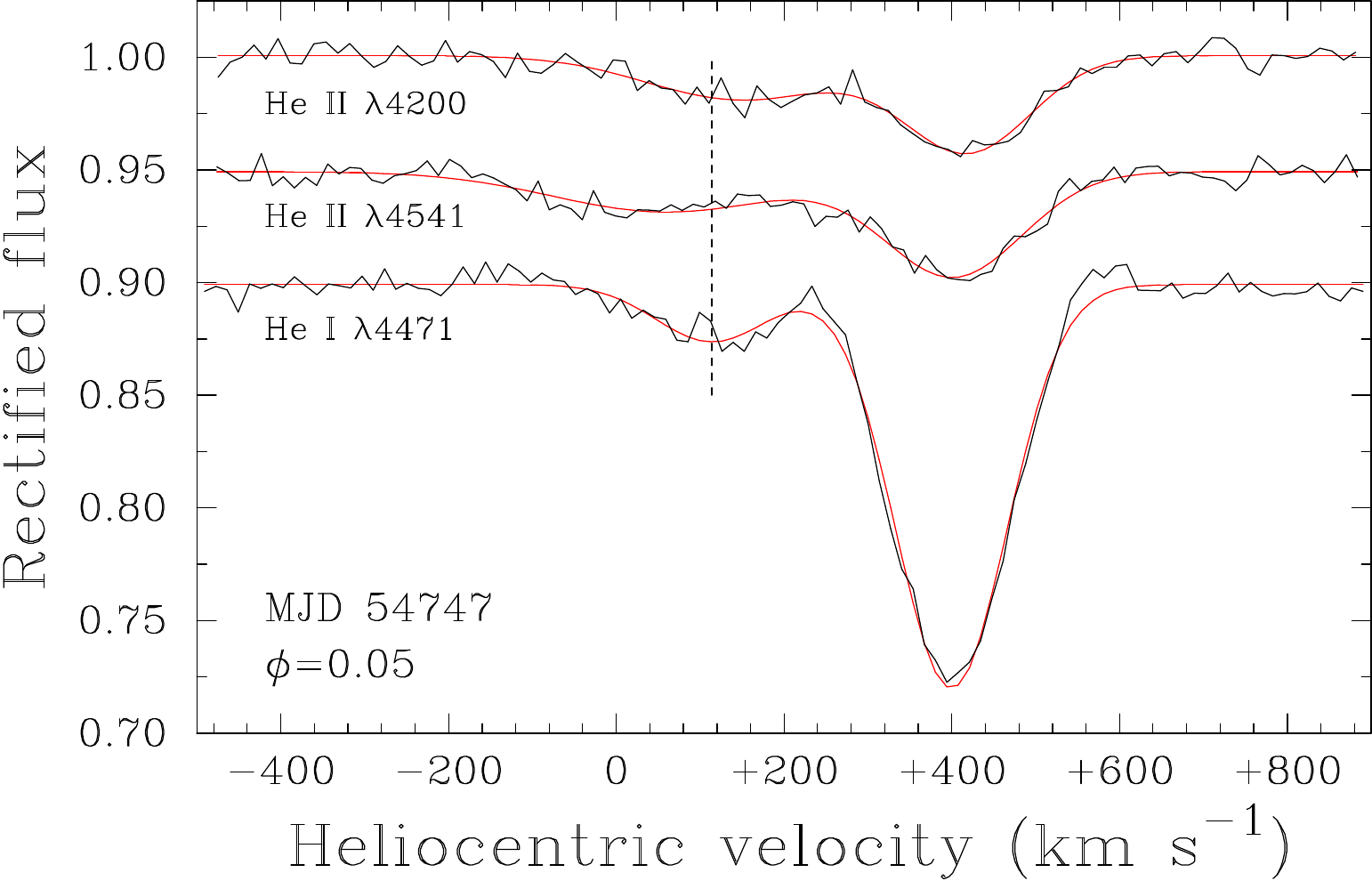} 
\caption{Selected helium lines in the spectrum of VFTS~450 near
  quadrature; smooth red curves show gaussian
  fits to the data.
  Although these helium lines give consistent
  results for the primary (at $\sim$+400~\kms),
  the secondary velocities are discordant
  (Section~\ref{v450:radvel}; for reference, the dashed vertical line
  indicates the measured secondary \hei~$\lambda$4471 velocity).
  This figure also illustrates the differences in \hei:\heii\ line
  ratios in the two components (Section~\ref{subsec:450}).}
\label{f_450_sb2}
\end{center}
\end{figure}

\begin{figure}
\begin{center}
\includegraphics[angle=0, scale=0.55]{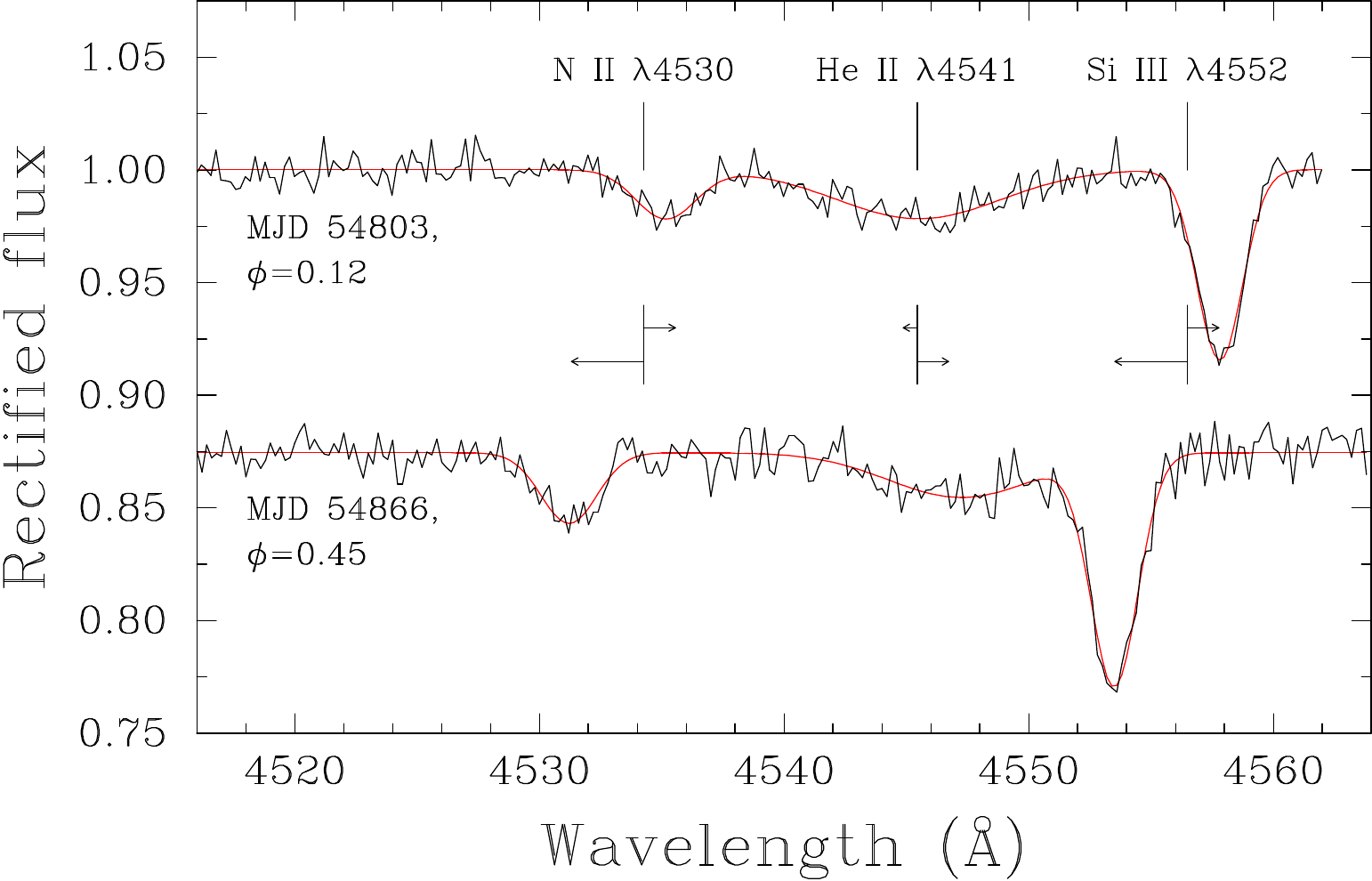} 
\caption{
VFTS~652 as a double-lined spectroscopic binary
(Section~\ref{s_652_para}).  Wavelengths of selected lines are shown
in the rest frame of the system centre of mass, together with the 
observed orbital
displacements of \nii~$\lambda$4530, \siiii~$\lambda$4552 (primary
spectrum), and \heii~$\lambda$4541 (secondary spectrum).}
\label{f_652_sb2}
\end{center}
\end{figure}

\paragraph{\emph{VFTS~450}}
\label{v450:radvel}

Because of blending with features in the primary spectrum, we did not
attempt radial-velocity measurements of the secondary at phases near
conjunctions ($100 \lesssim V_{\rm primary} \lesssim 400$~\kms).  The
helium lines in the secondary spectrum show poor agreement, as
illustrated in Fig.~\ref{f_450_sb2}; the \heii~$\lambda$4200
velocities are generally -- though not consistently -- some
$\sim$100~\kms\ more positive than found for
\ion{He}{ii}~$\lambda$4541 or \ion{He}{i} $\lambda$4471.  Given the
shallowness of the lines in the secondary spectra, we cannot rule out
that rectification difficulties contribute to this problem. In
practice, we rely principally on results for $\lambda$4200, which gives
consistent results and which is not subject to significant blending (cp.,
e.g., secondary $\lambda$4541, which can be affected by \ion{Si}{iii}
$\lambda$4552 in the primary spectrum).

\paragraph{\emph{VFTS~652}}

The absence of \heii\ lines in the primary spectrum renders measurement
of the weak \heii~$\lambda$4541 line in the secondary reasonably
straightforward in both LR02 and LR03 spectra. \heii~$\lambda$4200
gives consistent, but somewhat more scattered, results.

\begin{table*}
\centering
\caption{Radial-velocity orbital solutions, for circular and eccentric
  orbits.}  
\begin{tabular}{l|rr|rr}
\hline
\hline
\rule{0pt}{9pt}Parameter          & \multicolumn{2}{c|}{VFTS~450} &
\multicolumn{2}{c}{VFTS~652} \\
&\emph{circ.$\quad$}
&\emph{ecc.$\quad$}
&\emph{circ.$\quad$}
&\emph{ecc.$\quad$}\\
\hline
\rule{0pt}{9pt}$P_{\rm phot}$ (d)            &  \multicolumn{2}{c|}{6.892583}   &   \multicolumn{2}{c}{8.589555} \\
$\qquad\pm$                       &    \multicolumn{2}{c|}{0.000039}   &     \multicolumn{2}{c}{0.000090}     \\
$T_0^{\rm P}$ (MJD)            & 54761.421 &$\cdots\;\;$&54665.242&$\cdots\;\;$\\
$\qquad\pm$                      & 0.007 &&0.014\\
\rule{0pt}{12pt}$P_{\rm spec}$ (d)            
&  6.892325     &    6.892242   &   8.589534   &    8.589413      \\
$\qquad\pm$                       &    0.000124   &    0.000080   &   0.000089   &     0.000067     \\
$\gamma$ (\kms)                   &  248.46       &    246.26     &  253.09      &     255.32       \\
$\qquad\pm$                       &    1.61       &      1.10     &    0.83      &       0.65       \\
$K$ (\kms)                        &  208.8        &    207.6      &  201.6       &      199.8       \\
$\qquad\pm$                       &    2.2        &      1.5      &    1.4       &        1.0       \\
$e$                               & $\equiv$ 0    &     0.0722    &  $\equiv$0   &        0.0443    \\
$\qquad\pm$                       &               &     0.0071    &              &        0.0050    \\
$\Omega$ ($^\circ$)             &$\cdots\;\;$        &   356.2       & $\cdots\;\;$ &       10.0       \\
$\qquad\pm$                       &               &     6.3       &              &        5.4       \\
$T_0$ (MJD)                      &  54761.268     &  54761.214    &54897.050     &   54897.300      \\
$\qquad\pm$                       &    0.021      &      0.116    &    0.011     &       0.127      \\
$f(M)$ (\msun)                    &    6.51       &      6.35     &    7.31      &      7.10        \\
$\qquad\pm$                       &    0.20       &      0.13     &    0.15      &      0.11        \\
$a_1\sin{i}$ (\rsun)             &    28.43       &     28.20     &   34.22      &       33.88      \\
$\qquad\pm$                       &     0.29      &      0.20     &    0.24      &        0.18      \\
r.m.s. residual (\kms)            &    13.8       &       9.1     &     7.5      &        5.4       \\
\rule{0pt}{20pt}$q\,(=M_1/M_2)$  & \multicolumn{2}{c|}{$\qquad$0.61}& \multicolumn{2}{c}{$\qquad$0.40}\\
$\qquad\pm$        & \multicolumn{2}{c|}{$\qquad$0.05} & \multicolumn{2}{c}{$\qquad$0.05}\\
$M_1\sin^3(i)$ (\msun) 
                   & 10.4 & 10.2 & 5.6 & 5.5 \\
$\qquad\pm$        &  1.5 &  1.4 & 1.2 & 1.2 \\
$M_2\sin^3(i)$ (\msun) 
                   & 16.9 & 16.5 & 14.2 & 13.8 \\
$\qquad\pm$        &  1.2 &  1.0 &  1.1 &  1.1 \\
\hline
\end{tabular}
\tablefoot{Parameters are based on primary-star radial velocities,
  excepting the photometric period, $P_{\rm phot}$; the
  photometrically determined value of the time of circular-orbit
  maximum radial velocity, $T_0^{\rm P}$; and the mass ratio $q$
  (Section~\ref{sec:orb}).  Note that the $T_0$ parameter has
  different meanings for circular and eccentric orbits (times of
  maximum velocity and of periastron passage, respectively); for both
  targets, the
  numerical values from the spectroscopic solutions are coincidentally
  similar only because $\Omega \simeq 0^\circ$ in each case.}
\label{tab:specorb}
\end{table*}

%
%
%
%
%

\begin{figure}
\centering
\includegraphics[scale=0.35,angle=270]{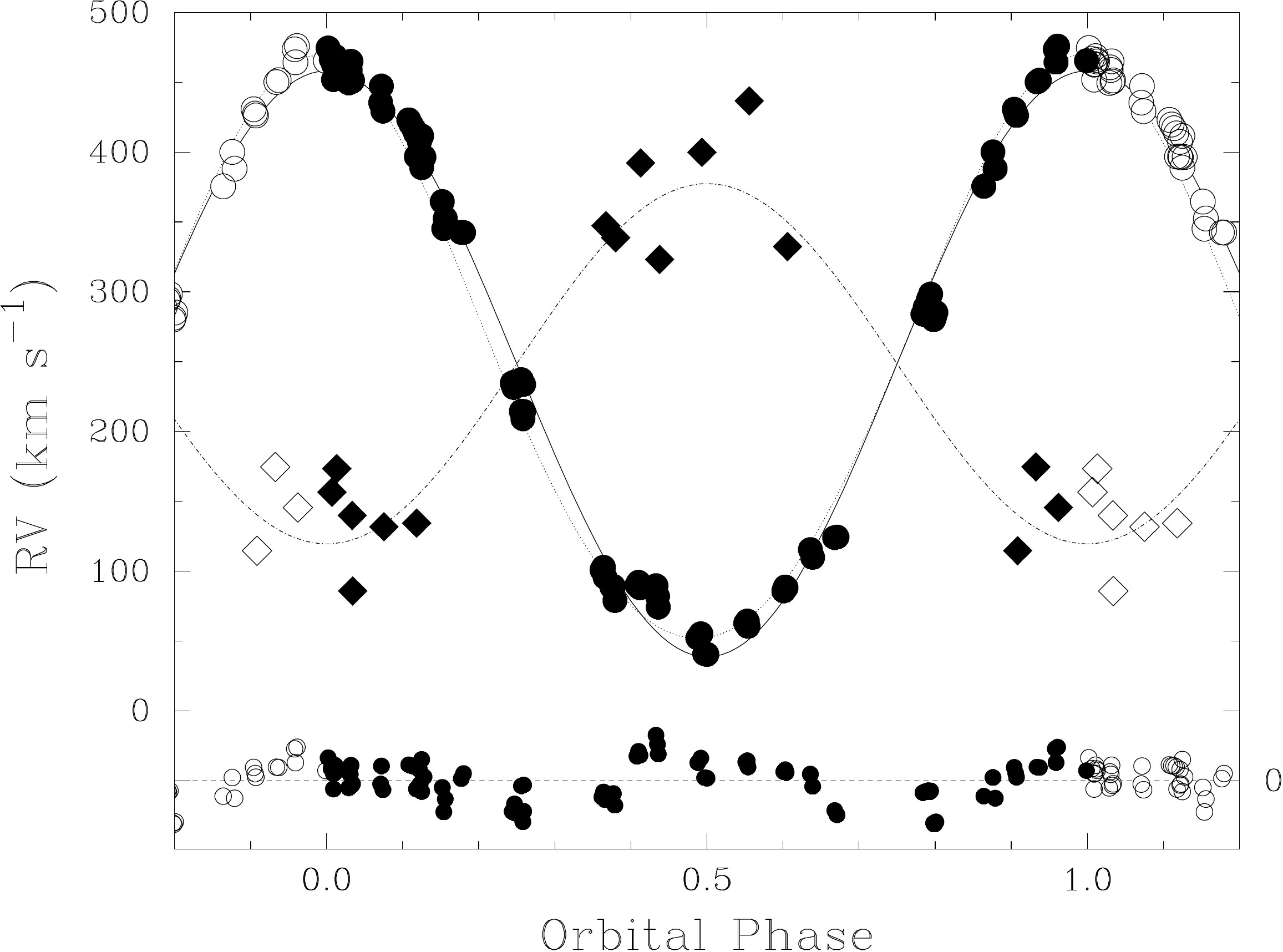}
\caption{Spectroscopic orbit for VFTS~450; orbital phases refer to the
  circular-orbit $T_0$ (from Table~\ref{tab:specorb}), as do the
  (O$-$C) residuals for the primary shown at the bottom of the plot.
  The eccentric-orbit solution for the primary is shown as a dotted
  line (which may appear as a continuous grey line if viewed at low
  resolution), and the circular orbit for the secondary as a dash-dot
  line.  Diamonds show the \heii~$\lambda${4200} velocities measured
  in the secondary's spectrum, and used to estimate the mass ratio.}
\label{fig:sb1}
\end{figure}

\begin{figure}
\centering
\includegraphics[scale=0.35,angle=270]{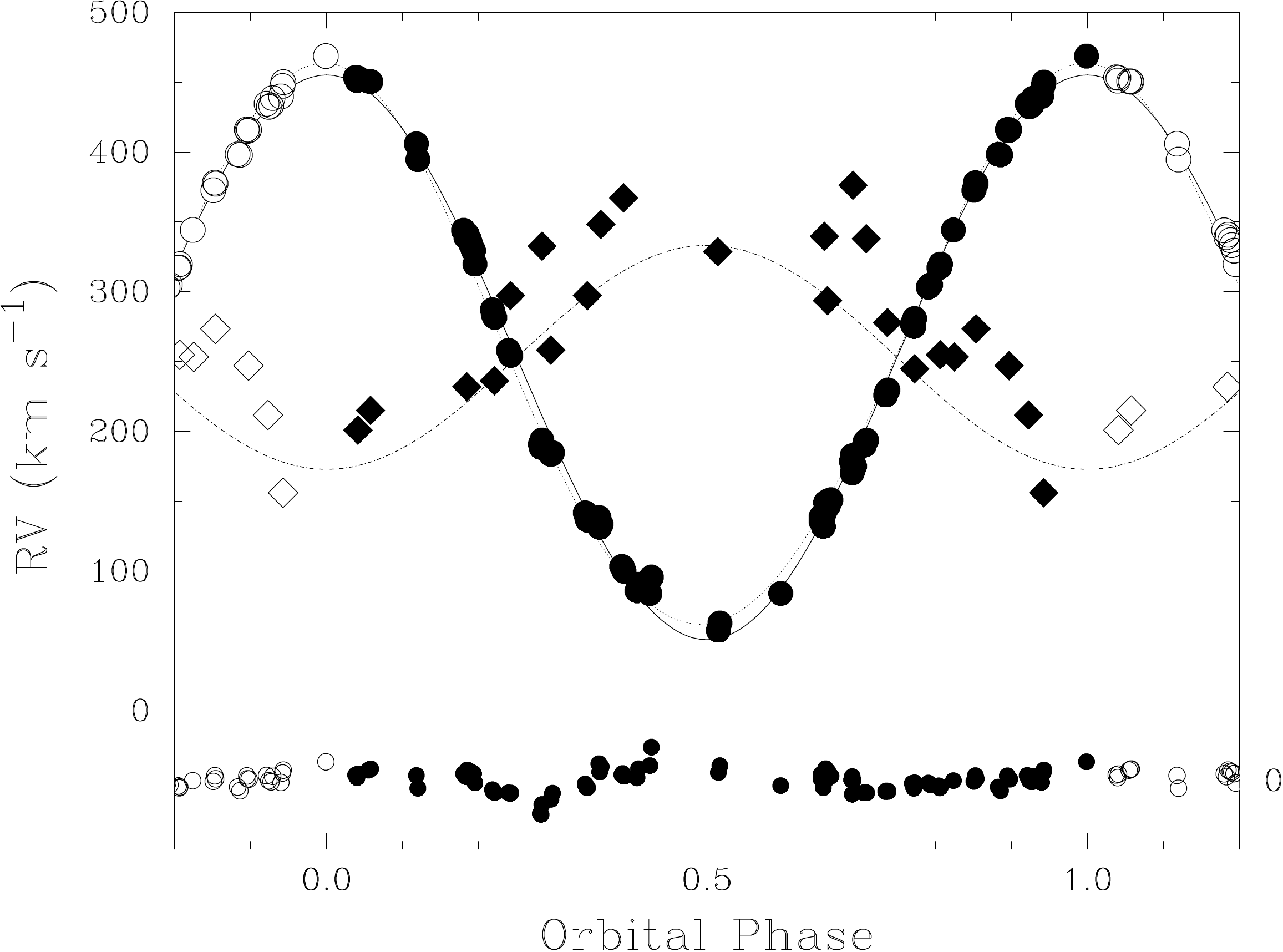}
\caption{Spectroscopic orbit for VFTS~652;  details as for
 Fig.~\ref{fig:sb1}, except that \heii~$\lambda$~4541 velocities are
 shown for the
 secondary.}
\label{fig:sb2}
\end{figure}

\begin{figure}
\centering
\includegraphics[scale=0.55,angle=0]{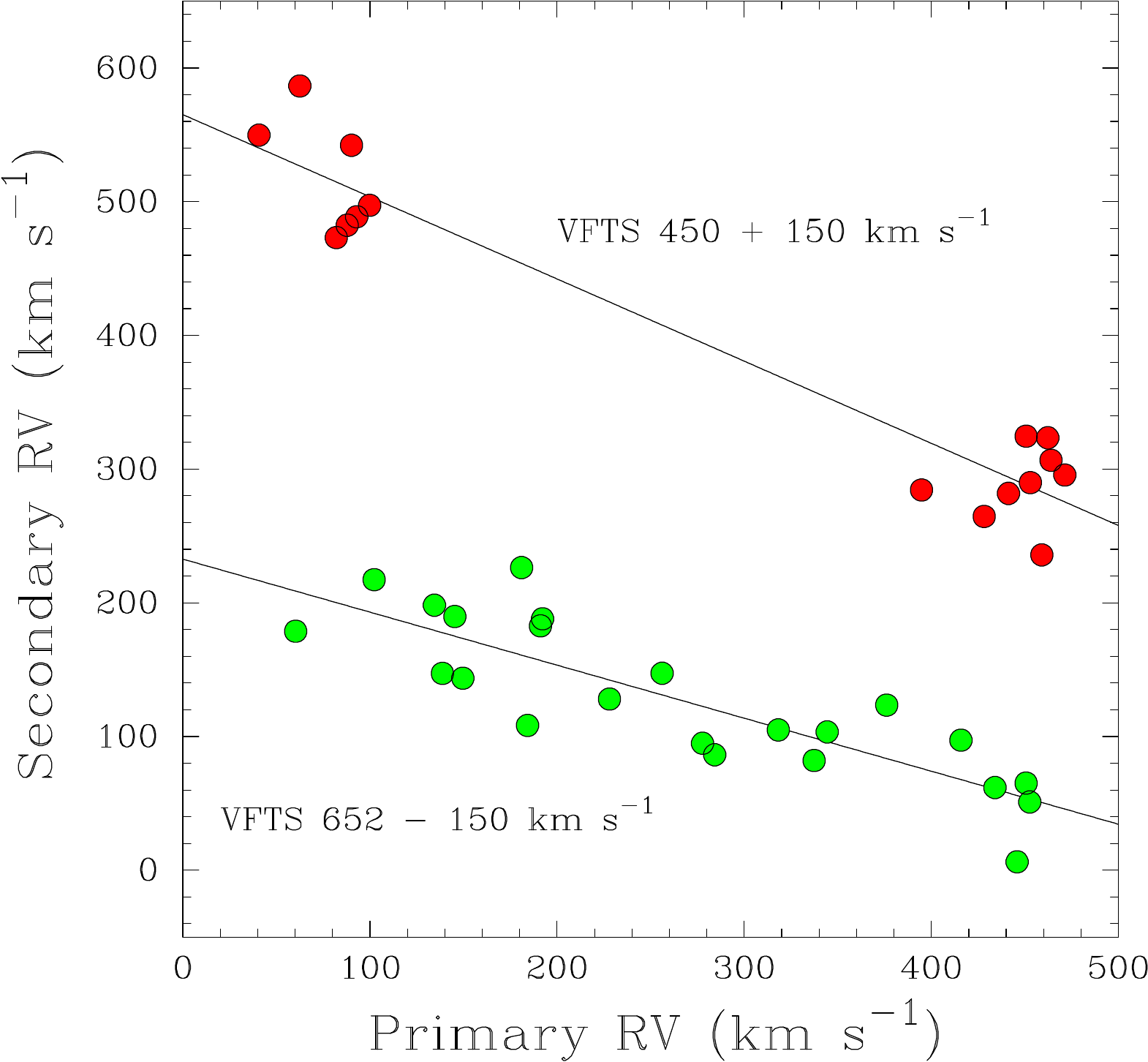}
\caption{Spectroscopic mass-ratio determination; secondary velocities
  have been offset as indicated for display purposes (primary
  velocities unchanged).   The slope of
  the linear fit gives the mass ratio directly.}
\label{fig:sb3}
\end{figure}

\begin{figure*}
\centering
\includegraphics[scale=0.65,angle=270]{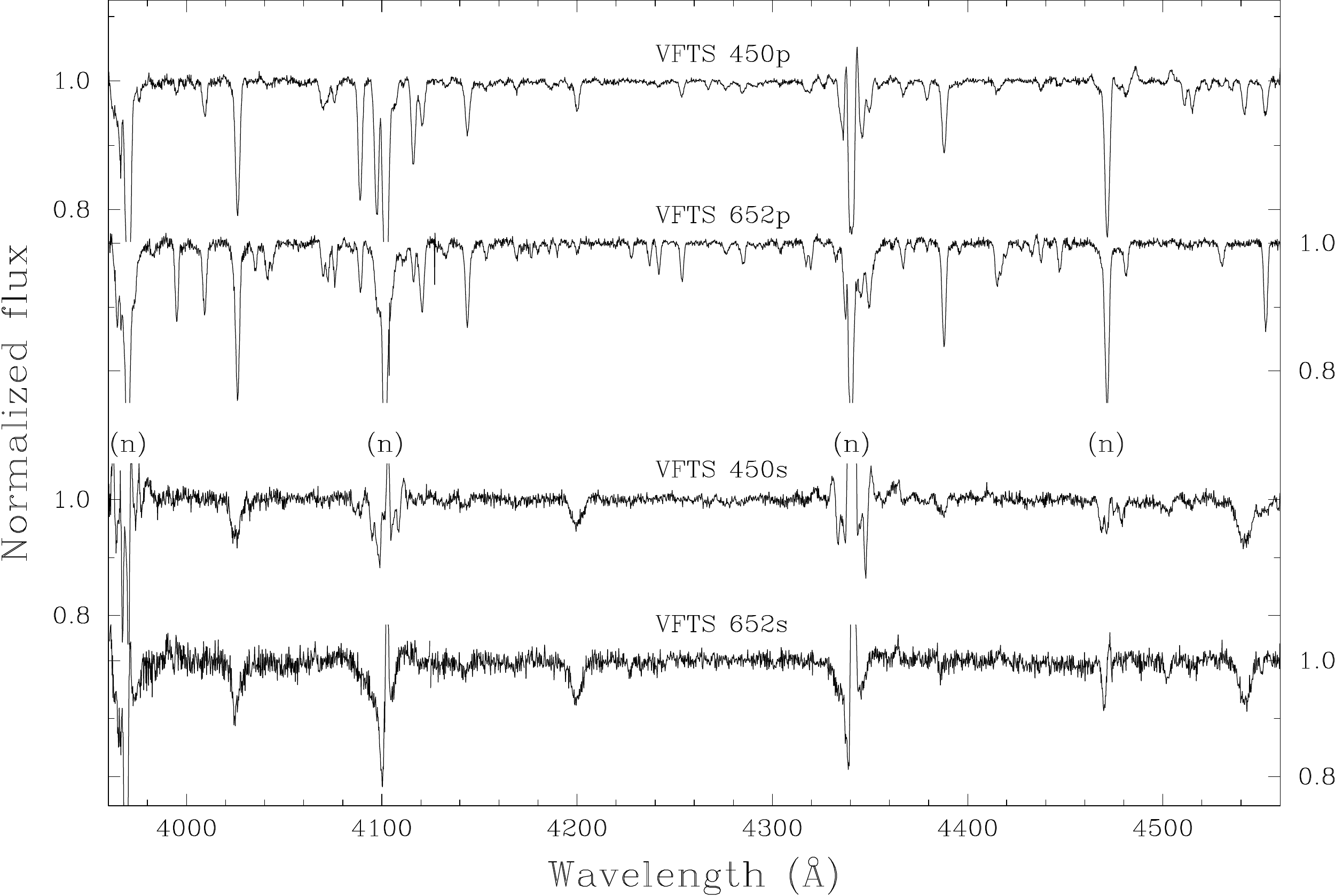}
\caption{Disentangled component spectra; the $y$-axis scaling is
  approximate (and depends on the primary:secondary continuum flux
  ratios).  The Balmer lines and, to a lesser extent, \hei\ are
  corrupted by nebular emission (marked as `{\tt n}'); broad features, such as
 the $\lambda$4430 diffuse
  interstellar band, are rectified out.}
\label{fig:dis}
\end{figure*}

\subsection{Results}
\label{sec:orb}

The primaries' spectroscopic orbits are summarized in
Table~\ref{tab:specorb}, and are illustrated in Figs.~\ref{fig:sb1}
and~\ref{fig:sb2}.  
We adopted uniform
weighting for our final solutions, but other weighting schemes result in
unimportant changes to the orbital parameters.  According to the
formulation of the \mbox{$F$ test} described by \citet{Lucy71}, the orbital
\mbox{eccentricities} are formally significant with $>99$\%\ confidence.
However, the apparent eccentricities are quite small, and we caution
that they may not reflect the true centre-of-mass
motions.

Given the considerable uncertainties in the radial-velocity
measurements of the secondaries, we chose a simple but robust method
to estimate the mass ratio for each system, namely, a linear
regression of the secondary velocities on the primary values
(Fig.~\ref{fig:sb3}).  The gradient yields the mass ratio directly,
independently of all other parameters; results are included in
Table~\ref{tab:specorb}.   
The means of the observed minus 
predicted secondary orbital velocities (i.e., the differences between primary
and secondary $\gamma$ velocities) are $+10.0{\pm}8.5$ (s.e.) and
$+29.5{\pm}6.5$~\kms\ for VFTS~450 and~652, respectively; 
differences in $\gamma$ velocities such as that shown by VFTS~652 
have been found previously
in `windy' massive binaries (e.g., \citealt{niemela86, niemala94}).

Although the spectroscopic period determined for
VFTS~450 differs from the  photometric value by $\sim$1.7-$\sigma$,
we don't consider this to be significant evidence for period changes,
given that the OGLE-III and OGLE-IV datasets are in good mutual agreement,
and span the spectroscopic epochs.

\subsection{Disentangling}
\label{sec:disentangle}

In principle, an alternative approach to the spectroscopic-orbit
modelling is a simultaneous solution of the individual component
spectra and the orbital characteristics (`disentangling'; cf., e.g.,
\citealt{hadrava04}).  Because of the weakness of the secondary
spectra we instead chose the simpler option of reconstructing the
separate component spectra in the more extensive LR02 datasets from
the `known' velocities, using {\sc{cres}} \citep{Ilijic04}.  We
explored the consequences of using observed velocities or those
calculated from the orbital solution, including using mass ratios in
the range 0.5--1.0 when computing the secondary spectra.  We found the
results to be quite robust to these factors (the corollary being that
the technique cannot recover a precise mass ratio for these data).
The resulting spectra of individual components are shown in
Fig.~\ref{fig:dis}; they have better S:N than any individual spectrum,
but the $y$ scaling is arbitrary, and nebular emission contaminates
some key lines.

\section{Model-atmosphere analysis}
\label{s_stellar}

\subsection{Methodology}
\label{s_method}

Model-atmosphere analyses of both systems were performed using a grid
of line-blanketed non-LTE {\sc tlusty} models at LMC metallicity
(\citealt{hub88}, \citealt{hub95}, \citealt{hub98}; for more details
of the grid see \citealt{rya03}, \citealt{duf05}).  The analyses
assume that each component's spectrum can be reliably characterized by
a single set of atmospheric parameters, and that hydrostatic,
plane-parallel structures are appropriate.  Depending on the adequacy
or otherwise of these assumptions, the results may be subject to
significant (and largely unquantifiable) systematic errors, and should
therefore be interpreted with due caution.


For the primary spectra, the atmospheric parameters were estimated
from the \ion{Si}{iii} and \ion{Si}{iv} line strengths, together with
the \ion{H}{i} and \ion{He}{ii} profiles.  The lower quality of the
secondary spectra allowed only relatively rough estimates of
parameter values to be made, using the \ion{H}{i} and \ion{He}{ii}
lines; the micro\-turbulence was indeterminate (and unimportant)
for these lines, and we assumed appropriate values.

The analyses were based principally on the disentangled spectra, which
have better signal:noise ratios than any individual spectrum (and, of
course, should be free from blending), although cross-checks were made
against the directly observed spectra, particularly for the Balmer
lines, which suffer nebular contamination.  

A complication is the uncertainty in the relative flux contributions
of the two components in each system; without ancillary information,
it is impossible to distinguish between a strong continuum with weak
lines and a weak continuum with strong lines.  We addressed this issue
by supposing that the primary [secondary] contributes a
wavelength-independent fraction \fsone\ [$\fstwo, = 1 - \fsone$] of
the rectified continuum flux, and adjusted this fraction as necessary.

\begin{table}
\centering
\caption{Equivalent widths of the primary in the integrated
spectra;  cf.\ $\S$\ref{s_wand}.}
\label{t_abund_lines}
\begin{tabular}{lcccc}
\hline \hline
\multicolumn{1}{c}{\rule{0pt}{11pt}Species} & \rule{0pt}{8pt}
Wavelength & Region & \multicolumn{2}{c}{$W_\lambda$ (m\AA)} \\
        & (\AA) && [450] & [652] \\
\hline
\rule{0pt}{10pt}$\quad$\ion{N}{ii}   & 3995.0 & LR02 & {\pa}26   & 185     \\
$\quad$\ion{N}{ii}                   & 4227.7 & LR02 & $\cdots$  & {\pa}39 \\
$\quad$\ion{N}{ii}                   & 4447.0 & LR02 & $\cdots$  & {\pa}70 \\
$\quad$\ion{N}{ii}                   & 4601.5 & LR03 & $\cdots$  & 113     \\
$\quad$\ion{N}{ii}                   & 4607.2 & LR03 & $\cdots$  & 117     \\
$\quad$\ion{N}{ii}                   & 4613.9 & LR03 & $\cdots$  & {\pa}77 \\
$\quad$\ion{N}{ii}                   & 4621.4 & LR03 & $\cdots$  & {\pa}84 \\
$\quad$\ion{N}{ii}                   & 4630.5 & LR03 & {\pa}41   & 189     \\
$\quad$\ion{N}{ii}                   & 4774.2 & LR03 & $\cdots$  & {\pa}12 \\
$\quad$\ion{N}{ii}                   & 4788.1 & LR03 & $\cdots$  & {\pa}24 \\
$\quad$\ion{N}{ii}                   & 4803.3 & LR03 & $\cdots$  & {\pa}51 \\
$\quad$\ion{N}{ii}                   & 4994.4 & LR03 & $\cdots$  & {\pa}40 \\

\rule{0pt}{10pt}$\quad$\ion{O}{ii}   & 4185.4 & LR02 & $\cdots$  & {\pa}20 \\
$\quad$\ion{O}{ii}                   & 4317.0 & LR02 & $\cdots$  & {\pa}61 \\
$\quad$\ion{O}{ii}                   & 4319.6 & LR02 & $\cdots$  & {\pa}59 \\
$\quad$\ion{O}{ii}                   & 4366.9 & LR02 & $\cdots$  & {\pa}63 \\
$\quad$\ion{O}{ii}                   & 4393.9 & LR02 & $\cdots$  & {\pa}14 \\
$\quad$\ion{O}{ii}                   & 4414.9 & LR02 & {\pa}26   & 102     \\
$\quad$\ion{O}{ii}                   & 4417.0 & LR02 & {\pa}16   & {\pa}72 \\
$\quad$\ion{O}{ii}                   & 4452.4 & LR02 & $\cdots$  & {\pa}12 \\
$\quad$\ion{O}{ii}                   & 4591.0 & LR03 & {\pa}24   & {\pa}62 \\
$\quad$\ion{O}{ii}                   & 4596.0 & LR03 & $\cdots$  & {\pa}44 \\
$\quad$\ion{O}{ii}                   & 4661.6 & LR03 & {\pa}34   & {\pa}96 \\
\rule{0pt}{10pt}$\quad$\ion{Mg}{ii}  & 4481.0 & LR02 & {\pa}40   & {\pa}70 \\
\rule{0pt}{10pt}$\quad$\ion{Si}{iii} & 4552.6 & LR02 & {\pa}98   & 230     \\
$\quad$\ion{Si}{iii}                 & \rq\rq & LR03 & {\pa}84   & 245     \\
$\quad$\ion{Si}{iii}                 & 4567.8 & LR03 & {\pa}69   & 198     \\
$\quad$\ion{Si}{iii}                 & 4574.8 & LR03 & {\pa}39   & 116     \\
\rule{0pt}{10pt}$\quad$\ion{Si}{iv}  & 4088.9 & LR02 & 322       & 104     \\
$\quad$\ion{Si}{iv}                  & 4116.1 & LR02 & 233       & {\pa}84 \\
\hline
\end{tabular}
\end{table}


\subsubsection{Metal-line equivalent widths}
\label{s_wand}

The data quality allows abundance analyses to be conducted for the
primary spectra. For this purpose, equivalent widths were measured by
fitting theoretical profiles to the observations, using a
least-squares technique.  In the LR03 region directly observed spectra
were used; for the LR02 setting, the disentangled spectra were employed
(to take advantage of the improved S/N), and the results scaled to
recover the $W_\lambda$ values that would be determined in the directly observed
spectra.  Results are summarized in Table~\ref{t_abund_lines}.  

The $\lambda$4552 \ion{Si}{iii} line falls in the region of overlap
between LR02 and LR03 spectra; results from both spectrograph
configurations are separately listed in the Table.  The values from
the two settings differ by $\sim$15m\AA, in opposite senses for
VFTS~450 and VFTS~652.  This is probably a fair reflection of
observational uncertainties (including, potentially, temporal/orbital variations,
although comparison of LR02 spectra taken at different epochs shows no
evidence for substantial changes in line strengths).

The disentangling results show that the absorption lines of metals can
be safely attributed to the primary spectra (even in the LR03 data),
but of course the measurements in Table~\ref{t_abund_lines} have to be
scaled by the appropriate {\ensuremath{\mathcal{F}}} value when performing an abundance
analysis.


\subsection{VFTS~450}
\label{subsec:450}


The LR02 primary spectrum shows hydrogen and neutral \& ionized
helium lines, together with strong metal lines (particularly 
\ion{O}{ii} and \ion{Si}{iv}). By contrast, only the hydrogen and
ionized helium lines are clearly seen in the secondary spectrum, with
the former being badly contaminated by nebular emission. The
\ion{He}{i} lines at 4387, 4471\AA\ are probably present, with less
convincing evidence for \ion{Si}{iv} $\lambda$4089 and
\ion{N}{iii} $\lambda\lambda$4097, 4510--4534.

\begin{table*}
\caption{Summary of stellar-atmosphere parameter determinations.}
\begin{center}
\begin{tabular}{lcccccccccccc}
\hline\hline
\multicolumn{1}{c}{Star}  & \rule{0pt}{8pt}\fsone\ & \fstwo\ & \teff & \logg & $\xi$ & {\vesini}
& \multicolumn{5}{c}{Abundances} \T\\
& & & (kK) & (cgs) &(\kms) &(\kms) & N & O & Mg & Si &N/O\\
\hline
\rule{0pt}{9pt}{\ps}VFTS~450  p    & 1.00 & 0.00 & 25.5 & 2.7 & 15 & {\pa}99 & 7.32 & 7.75 & 6.93 & 6.70  &$-0.43$\\
$\ast$VFTS~450      p   & 0.75 & 0.25 &  27.0 & 2.9 & 15 &  {\pa}99 & 7.63 & 8.06 & 7.13 & 7.17  &$-0.43$\\
{\ps}VFTS~450                 p    & 0.50 & 0.50 & 29.0 & 3.1 & 18 &  {\pa}99 & 8.02  & 8.48 & 7.34 & 7.68 &$-0.46$   \\
\rule{0pt}{9pt}$\ast$VFTS
     450                 s   & 0.75 & 0.25 & 33.5:35.0 & $\le$3.8\px & [10] & 320 & $\cdots$ & $\cdots$ & $\cdots$ & $\cdots$ & $\cdots$ \\
{\ps}VFTS~450                 s    & 0.50 & 0.50 & 28.5:31.0 & $\le$3.4\px & [10] & 320 & $\cdots$ & $\cdots$ & $\cdots$ & $\cdots$ & $\cdots$ \\
{\ps}VFTS~450                 s    & 0.00 & 1.00 & 25.5:27.5 & $\le$3.0\px & [10] & 320 & $\cdots$ & $\cdots$ & $\cdots$ & $\cdots$ & $\cdots$ \\
\hline                                                                                                       
\rule{0pt}{9pt}{\ps}VFTS~652  p    & 1.00 & 0.00 & 21.3 & 2.6 & {\pa\pb}9\pb &{\pa}83 &   8.04    & 7.88     & 6.89      & 6.94  &+0.16\\
$\ast$VFTS~652      p   & 0.75 & 0.25 & 22.2 & 2.8 & \pb13\pb& {\pa}83 & 8.13      &  7.96      & 7.00      & 7.07   &+0.17\\
{\ps}VFTS~652                 p    & 0.50 & 0.50 & 23.0 & 3.0 & \pb19\pb& {\pa}83  & 8.31      & 8.08     & 7.15      & 7.28 &+0.23     \\
\rule{0pt}{9pt}$\ast$VFTS 
     652                 s   & 0.75 & 0.25 & 35.0 & 3.7 & \pa[5]  & 260 & $\cdots$ & $\cdots$ & $\cdots$ & $\cdots$ & $\cdots$ \\
{\ps}VFTS~652                 s    & 0.50 & 0.50 & 30.0 & 3.1 & [10] & 260 & $\cdots$ & $\cdots$ & $\cdots$ & $\cdots$ & $\cdots$ \\
{\ps}VFTS 652                 s    & 0.00 & 1.00 & 26.0 & 2.7 & [10] & 260 & $\cdots$ & $\cdots$ & $\cdots$ & $\cdots$ & $\cdots$ \\
\hline
\multicolumn{7}{r}{Reference LMC baseline abundance:}   &\rule{0pt}{8pt}6.90 & 8.35 & 7.05 & 7.20 &$-1.45$   \\
\hline
\end{tabular}
\tablefoot{$\mathcal{F}_{1,2}$
are the adopted fractional contributions of the primary, secondary (p, s)
components to the total
continuum flux, and $\xi$ is the microturbulence (with assumed values
given in square brackets).
Baseline LMC abundances 
are taken from 
\citeauthor{hun07}
(\citeyear{hun07}, on a logarithmic scale where the abundance
of hydrogen by number  $\equiv{12}$).  Abundances are given to two
decimal places to clarify the sensitivity to $\mathcal{F}$ values, and not to indicate
the accuracy of the determinations (for which realistic errors are
$\gtrsim{0.2}$~dex, as discussed in Section~\ref{sec:reliar}).
Similarly, projected rotation velocities are probably good to only
$\sim$10--20\%\
($\S$\ref{sec:vsini}).
Preferred solutions are indicated by asterisks.}
\end{center}
\label{t_comparison}
\end{table*}

\paragraph{\emph{Primary:}}

The primary spectrum was first analysed by assuming no secondary
contamination (i.e., $\fstwo \equiv 0$), leading to estimates of
\teff\ and \logg\ that are, in practice, lower limits to allowable
values.  To investigate the sensitivity of the results to spectral
contamination by the secondary component, the analysis was repeated
for $\fstwo = 0.25$, 0.5, spanning the range of plausible values;
results are summarized in Table \ref{t_comparison}.

Gravities were estimated from the H$\delta$ and H$\gamma$
line profiles, with results agreeing to better than 0.1~dex.  The
effective temperature was estimated from the \ion{He}{ii} spectrum (by
assuming a normal helium abundance), as results from the silicon
ionization equilibrium were found to be sensitive to the
micro\-turbulence, $\xi$, and were used to determine that parameter.
(Because of the relatively large value of $\xi$, the \ion{Si}{iii}
triplet lines at 4552--4574\AA\ lie near the linear part of the curve
of growth, and hence are not particularly sensitive to the
micro\-turbulence.)  

Given $\xi$ and {\ensuremath{\mathcal{F}}}, element abundances were
estimated from the equivalent widths listed in
Table~\ref{t_abund_lines}, with standard deviations of 0.1--0.2~dex
implied by the individual oxygen estimates.
To obtain
approximately `normal' LMC abundances for magnesium and silicon requires
$\fstwo \simeq 0.25$, which represents our `best-bet' model.  There
then appears to be a significant surface-nitrogen enhancement
approaching 1~dex, and an oxygen depletion of $\sim$0.3~dex.

The range of parameter estimates, together with the agreement between
theoretical and observed profiles, leads us to adopt modelling
uncertainties of $\pm$1~kK in \teff, $\pm$0.2~dex in \logg, and
2~\kms\ in $\xi$.  Uncertainties on the abundances are difficult to
address precisely; the adopted uncertainties in the atmospheric
parameters alone translate into typical errors of 0.15~dex for both
nitrogen and oxygen, (see \citealt{hun07} for more details). Varying
the secondary contribution (\fstwo) contributes significant additional
uncertainty to the absolute abundances, but has little effect on the N:O
abundance ratio;  the inference of a significant surface-nitrogen
enhancement appears to be robust.

\paragraph{\emph{Secondary:}}

The weakness of the secondary's absorption lines makes an atmospheric
analysis difficult, and our results should be treated with caution.
None\-the\-less, the \ion{He}{ii} absorption lines do provide useful
diagnostics, principally for the effective temperature.  Additionally,
although the Balmer-series lines are badly contaminated by nebular
emission, the lack of significant Stark-broadened wings sets an upper
limit on the surface gravity. As an exercise in defining the range of
possible parameter space, we conducted analyses assuming that the
continuum was entirely due to the secondary ($\fstwo = 1$), along with
two cases considered for the primary ($\fstwo$ = 0.25 and 0.5); the
results are summarized in Table~\ref{t_comparison}. Note that the
gravity limits are appropriate for the upper limit of the
effective-temperature range -- a lower effective temperature would
lead to lower gravity limit.

\subsection{VFTS~652}
\label{s_652_para}

The spectra of VFTS~652 show a rich metal-line spectrum for the
primary, in accord with its classification as a B-type
super\-giant. The secondary spectrum shows convincing evidence for the
presence of \ion{He}{ii} lines (Fig.~\ref{f_652_sb2}), together with
Stark-broadened wings in the Balmer series; \ion{He}{i} lines
also appear to be present.

\paragraph{\emph{Primary:}}

As for VFTS~450, we evaluated parameters for $\fstwo = 0$, 0.25, and 0.5.
The effective
temperature was estimated from the silicon ionization equilibrium, the
gravity from the Balmer-line profiles, and the micro\-turbulence from
the relative strengths of lines in the \ion{Si}{iii} triplet. 
Results are summarized in Table~\ref{t_comparison}.

To obtain `normal' LMC abundances for magnesium and silicon requires
$\fstwo=0.25$--0.50. Abundances were also derived for nitrogen and
oxygen, with the scatter among estimates from individual lines being
in the range 0.1--0.2~dex. Using the same criteria as for VFTS~450
leads to estimated uncertainties of $\pm$1~kK, $\pm$0.1~dex, and
$\pm$3~\kms, for \teff, \logg, and $\xi$ respectively. These in turn
imply uncertainties of typically 0.2--0.3~dex for the nitrogen and
oxygen abundances (but significantly less for the N:O abundance
ratio). The effects of varying the dilution factor are relatively
small, leading to adopted final errors of 0.3~dex for these elements;
a conservative error estimate 
on the N:O ratio is $\sim$0.2~dex.

At $\fstwo = 0.25$, the oxygen abundance is possibly under\-abundant,
by $\sim$0.4~dex compared to the LMC baseline, while nitrogen is again
clearly enhanced, by more than 1.0~dex. Qualitatively similar
conclusions follow for other dilution factors; regardless of the
dilution factor adopted, the N:O abundance ratio is 1.6--1.7~dex
higher than for the adopted LMC baseline abundances.

\paragraph{\emph{Secondary:}}

Again as an exercise in defining the range of possible parameter
space, we used the \ion{He}{ii} $\lambda\lambda$4200,~4541 and
available Balmer lines to estimate atmospheric parameters for several
dilution factors. The results are summarized in Table
\ref{t_comparison}; for $\fstwo = 0.25$, inferred stellar parameters
lie at the boundary of, or just outside, our grid of {\sc tlusty}
models. Representative values of the micro\-turbulence were adopted
but varying these by reasonable amounts would have a negligible
effect on our estimates.

The absence of detectable metal lines in the secondary spectrum is
consistent with the adopted parameters, secondary flux level, and
\mbox{signal:noise} ratio.

\subsection{Uncertainties}
\label{sec:reliar}


\paragraph{\emph{Atmospheric parameters:}} 
The primary spectrum of VFTS~652 was easier to model than that of
VFTS~450, with better internal agreement between different spectral
features.  However, for both systems the inclusion of dilution by the
secondary leads to only relatively small changes in fit parameters. We
therefore expect the \mbox{error} \mbox{estimates} for the primaries'
\teff\ and \logg\ values discussed above 
to be reasonable.

Parameter estimates for the secondary components are considerably less
secure. The \teff\ and \logg\ values estimated from the
\ion{He}{ii} and \ion{H}{i} profiles are moderately sensitive to
the choice of the dilution factor, \fstwo. It is therefore difficult
to assess realistic error estimates, although in general terms we
consider the \teff\ values to be more reliable than those for
\logg\ (reflecting the greater reliability of the deconvolved
\ion{He}{ii} spectrum); for the `worst case' of VFTS~450, we estimate
an uncertainty in \teff\ of perhaps $\pm$4~kK. Nonetheless,
provided that the secondary
spectra are formed in the photo\-spheres of the secondary stars, it
seems secure that the primary is the
cooler component in each system.

\paragraph{\emph{Dilution factor:}} From the general
characteristics of the spectra, we are confident that \fstwo\ is
certainly less than 0.5 for each system. Unfortunately, the magnesium
and silicon abundances used to constrain on the dilution factor also
depend the atmospheric parameters (and hence do not provide
particularly strong limits). However, that the secondary spectra can
be measured at all implies $\fstwo \gtrsim 0.1$. Hence dilution
factors of $\fstwo \simeq 0.25\pm 0.1$ would appear to be
reasonable for both systems.

\paragraph{\emph{Abundance estimates:}} These are available
only for the primary components, and are quoted in Table
\ref{t_comparison} to two decimal places in order to illustrate the
sensitivity to the adopted dilution factors (and \emph{not} to indicate their
accuracy). The most striking results are the enhanced
surface-nitrogen abundances (0.7\,dex for VFTS~450, 1.2\,dex for
VFTS~652) and nitrogen:oxygen abundance ratios (1.0\,dex and 1.6\,
respectively); although there are additional uncertainties associated with
binarity, the relatively normal abundance estimates for other elements
strongly support large nitrogen enhancements in both stars. 

The situation is less clear for oxygen, with implied underabundances
of 0.3--0.5~dex, compared to uncertainties of $\pm$0.2--0.3~dex; such
underabundances are, however, consistent with those predicted from LMC
single-star evolutionary models that yield nitrogen enhancements of a
factor $\sim$10 \citep[e.g.,][]{bro11a}.

\subsection{Projected rotation velocities}
\label{sec:vsini}

Estimates of the projected equatorial rotation velocities, \vesini,
were obtained by using a Fourier technique similar to that adopted in
other VFTS rotational-velocity investigations \citep[see, for
  example][]{duf12, ram13}, and by simple profile fitting, which
yields the line-width parameter \vsini\ (which has contributions from
both rotation and macroturbulence).  Measurements were principally
made on the disentangled spectra ($\S$\ref{sec:disentangle}), but
checks were performed using the directly observed data.

For the primaries, we used \ion{N}{ii}~$\lambda\lambda$3995, 4447,
\ion{Si}{iii}~$\lambda$4552, \ion{Si}{iv}~$\lambda$4089, and
\ion{S}{iii}~$\lambda$4253.  Adopted \vesini\ values are averages of
results from all lines in each target; although the dispersion from
different lines is $<$5\%, systematic effects can be important
\citep[see notes in][]{sun13,sim14}, and realistic uncertainties are
perhaps $\sim$10--20\%, following arguments given by
\citet{mce15}.  For the secondaries we were limited to the available
\heii\ lines (Section~\ref{s_stellar}), but systematic effects should
be negligible, and likely uncertainties are on the order of $\sim\pm$10\%.

Results are included in Table~\ref{t_comparison}.  The primaries'
rotation velocities are rather high when considered in the context of
the VFTS sample of single late-O/early-B super\-giants \citep{mce15},
but are close to values expected for synchronous rotation
($\S$\ref{s:syspar}); however, the secondaries' rotations are
exceptionally rapid.  

Estimates of \vsini\ from simple profile fitting of the primary
spectra are $105 \pm 6$ and $91 \pm 4$~\kms\ (VFTS~450, 652),
$\sim$5--10\%\ larger than \vesini\ measurements from
the Fourier Transform methodology; however, the estimates are
consistent within the uncertainties. Thus while there may be a
macroturbulent contribution to line broadening in the primary spectra,
its extent is difficult to quantify usefully.  Rotational
broadening dominates the secondary spectra, rendering estimates of any
macroturbulence contribution impossible.

\section{Discussion}
\label{sec:sys}

The secondary-spectrum radial velocities indicate that the
spectroscopically less conspicuous star is, in each system, the more
massive component.  This conclusion is supported by photometric
considerations ($\S\S$\ref{sec:450con}, \ref{sec:652con}), and so appears
to be a robust conclusion, even if the secondary spectrum is only an
approximate tracer of the secondary star's centre-of-mass motions.

Taken at face value, the spectroscopic analysis also indicates the
secondaries to be the hotter components and hence, being fainter, the
smaller.  However, in each system, the secondary spectrum has the
larger \vesini; we cannot, therefore, assume corotation in order to
constrain radii or inclinations.  In principle, a light-curve analysis
can yield this information, but eclipses, if they occur at all, are
very shallow (Fig.~\ref{fig:photom}), leading to poorly constrained
solutions, and our initial attempts in this direction have yielded
unphysical results.  The lack of deep eclipses, coupled with
significant `ellipsoidal' variations, does, though, indicate both
that the primaries are close to filling their Roche lobes, and that
the systems are observed at intermediate orbital inclinations.

\subsection{System constraints}
\label{sec:syscon}

The systems' absolute magnitudes can be determined from the apparent
magnitudes (we use the mean $V, I_{\rm C}$  OGLE results from
Table~\ref{t_photo}), the LMC distance modulus (18.5;  cf.,
e.g., \citealt{Schaefer08}, \citealt{Pietr13}), and the reddenings.

We estimate $(B-V)_0$ and $(V-I_{\rm C})_0$ by taking flux-weighted
averages of empirical intrinsic colours for each component as a
function of spectral type (from \citealt{wegner94}), and of model
colours as a function of temperature (making use of synthetic
photometry from the LMC-abundance {\sc Atlas} models reported by
\citealt{Howarth11a}).  The two sources of intrinsic colours, with two
observed colours (Table~\ref{t_photo}), and a reddening law
(\citealt{howarth83}; $A(I_{\rm C})/E(B-V)=1.84$), then yield four
separate estimates of $E(B-V)$ and $E(V-I_{\rm C})$, whence four
estimates of $M(V)$ and $M(I_{\rm C})$ for each system.  The
dispersions in these estimates are small, and we simply adopt mean
values; for quantitative results we rely principally on the $I_{\rm
  C}$-band results, since the extinction and the sensitivity of flux
to temperature are both slightly less here than at $V$.  (Adopting the
absolute $V$ magnitudes introduces only minor changes to the numerical
results, as illustrated in Fig.~\ref{f_masslum}.)

Absolute magnitudes for the individual components follow from
the continuum flux ratio, \fstwo/\fsone [$\equiv\fstwo/(1-\fstwo)$]. Coupling these with the corresponding surface
fluxes (from model atmospheres at the spectroscopically-determined
effective temperatures) gives the stellar radii.

The observed $a_1\sin{i}$ value, together with the mass
ratio $q$, gives both the projected semi-major axis $a\sin{i}$, and
the projected Roche-lobe radii, $R_{\rm L}(1,2)\sin{i}$ (conveniently
evaluated using the analytical approximation given by
\citealt{Eggleton83}).  Requiring the primary's radius not to exceed
its Roche-lobe radius sets a limit on 
$\sin{i}$ for a given $q$;  or, alternatively, limits
possible values for $q$ (by setting $\sin{i} = 1$).

With values for $R$, \teff, $q$, and $i$ in hand, other para\-meters
($L$, $M$, etc.) follow straightforwardly, given the primary's spectro\-scopic
orbit.

\subsubsection{VFTS~450}
\label{sec:450con}

We find $M(V) = -6.35 \pm 0.11$, $M(I_{\rm C}) = -6.01\pm 0.07$, where the
errors are standard deviations of the four individual estimates
(which are not independent).  The inferred reddening is slightly
larger for the
$(B-V)$ baseline [$E(B-V)=0.45 \rightarrow E(V-I_{\rm C})=0.57$] than
it is for $(V-I_{\rm C})$
[$E(B-V)=0.39 \leftarrow E(V-I_{\rm C})= 0.49$].

The upper limit on the primary radius, assuming that it contributes
\emph{all} the $I_{\rm C}$-band light, is
{25.4}~\rsun, for $\teff(1) = 27$~kK $(\pm{0.8}$, $\pm{1.1}$\rsun\ for
$\Delta\teff\ = \mp1$~kK, $\Delta{M} = \mp0.1$).  More realistically,
using the spectro\-scopic \mbox{($\sim{B}$-band}\footnote{For
    completeness, we adjust the \fstwo\ values as a function of
    wavelength by using model-atmosphere fluxes, though this has
    negligible consequences.}) brightness ratio of
$\sim$3:1, the implied radii are $R_{1,2} \simeq 22.0$,  $10.1$~\rsun,
with uncertainties on the order of 10\%.

For a primary radius 
$R_1 \le 25.4 \pm 1.0$~\rsun\ we find $q \le 1.24 \pm 0.06$.  This can
be considered a rather firm upper limit, as it depends only on the
absolute magnitude, the primary's effective temperature, and its
radial-velocity curve, all of which are reasonably well established;
this analysis therefore suggests that the secondary is very probably the more
massive component (independently of the secondary radial-velocity
curve).  Adopting the spectroscopic mass ratio of 0.61
implies $\sin{i} \le 0.70$, where the equality corresponds to a
lobe-filling primary.

%
%
%
%

%

\subsubsection{VFTS~652}
\label{sec:652con}

We estimate $M(V)= -5.86 \pm 0.14$, $M(I_{\rm C}) = -5.67 \pm 0.08$;
in this case,
the inferred reddening is slightly
smaller for the
$(B-V)$ baseline [$E(B-V)=0.40 \rightarrow E(V-I_{\rm C})=0.50$] than
for $(V-I_{\rm C})$
[$E(B-V)=0.47 \leftarrow E(V-I_{\rm C})= 0.59$].

The same reasoning as applied in $\S$\ref{sec:450con} gives an upper
limit on the primary's radius of $R_1 \le
25.5$\rsun\ ($\pm{0.9}$, $\pm{1.2}$\rsun).  This implies $q \le
0.98\pm0.05$ (again indicating that the secondary is the more massive
component), or, adopting the spectroscopic mass ratio, $\sin{i} \le
0.66$; while the spectroscopic brightness ratio of $\sim$3:1 implies
$R_{1,2} \simeq 22.1, 8.5$~\rsun.

\begin{sidewaystable*}
\caption{System parameters based on spectroscopy.}
\centering
\begin{tabular}{lrccccccrcccccc}
\hline
\hline
\multicolumn{2}{l}{Parameter}&\multicolumn{5}{c}{VFTS 450} &$\,   $&& \multicolumn{5}{c}{VFTS 652} \T           \\ \cline{2-7}\cline{9-14}\\
             & M1 \B &$\Delta{M(I_{\rm C})}$  & $\Delta{q}$              & $\Delta\fstwo$           &  $\Delta\teff(1)$&$\Delta\teff(2)$&$\qquad$
             & M1    &$\Delta{M(I_{\rm C})}$  & $\Delta{q}$              & $\Delta\fstwo$           &  $\Delta\teff(1)$&$\Delta\teff(2)$\\
\hline
\rule{0pt}{9pt}$M(I_{\rm C})$
\T                 &  $-6.01$ &$\pm$0.20      &              &               &              &              &  &$-5.67$  & $\pm$0.20      &               &                &                &              \\
$q$                &   0.61   &               & $\pm$0.05    &               &              &              &  & 0.40    &                & $\pm$0.05     &                &                &              \\
\fstwo             &   0.25   &               &              & $\pm$0.10     &              &              &  & 0.25    &                &               & $\pm$0.10      &                &              \\
\teff(1)\ (kK) \T  &   27.00  &               &              &               & $\pm$1.0\pa  &              &  & 22.2    &                &               &                & $\pm$1.0\pa    &              \\
\teff(2)\ (kK)     &   34.25  &               &              &               &              & $\pm$2.0\pa  &  & 35.0    &                &               &                &                & $\pm$2.0\pa  \\
\\                                                                                                                                                                                            
$R_1/\rsun$        &   22.0   &$\mp$2.0\pa    &$\cdots\;\;$  &$\mp$1.5\pa    &$\mp$0.7\pa   &$\cdots\;\;$  &  & 22.1    &$\mp$2.0\pa     &$\cdots\;\;$   &$\mp$1.5\pa     &$\mp$0.8\pa     &$\cdots\;\;$  \\
$M_1/\msun$        &   29.7   &$\mp$8.4\pa    &$\mp$0.3\pa   &$\mp$5.9\pa    &$\mp$2.8\pa   &$\cdots\;\;$  &  & 20.1    &$\mp$5.7\pa     &$\mp$0.2\pa    &$\mp$4.1\pa     &$\mp$2.2\pa     &$\cdots\;\;$  \\
$\log(L_1/\lsun)$  &   5.36   &$\mp$0.08      &$\cdots\;\;$  &$\mp$0.06      & $\pm$0.04    &$\cdots\;\;$  &  & 5.03    &$\mp$0.08       &$\cdots\;\;$   &$\mp$0.06       & $\pm$0.05      &$\cdots\;\;$  \\
$\logg_1$ (cgs)    &   3.23   &$\mp$0.04      &$\mp$0.00     &$\mp$0.03      &$\mp$0.01     &$\cdots\;\;$  &  & 3.05    &$\mp$0.04       &$\mp$0.00      &$\mp$0.03       &$\mp$0.02       &$\cdots\;\;$  \\
\vcpsini\ (\kms)    &   113    &$\cdots\;\;$   & $\pm$6       &$\cdots\;\;$   &$\cdots\;\;$  &$\cdots\;\;$  &  & 85      &$\cdots\;\;$    & $\pm$6        &$\cdots\;\;$    &$\cdots\;\;$    &$\cdots\;\;$  \\
\\                                                                                                                                                                                                       
$R_2/\rsun$        &   10.1   &$\mp$0.9\pa    &$\cdots\;\;$  & $\pm$2.1\pa   &$\cdots\;\;$  &$\mp$0.5\pa   &  & 8.5     &$\mp$0.8\pa     &$\cdots\;\;$   & $\pm$1.7\pa    &$\cdots\;\;$    &$\mp$0.4\pa   \\
$R_2/R_{\rm L}$& 0.37 &$\cdots\;\;$   &$\pm$0.01     & $\pm$0.10     &$\pm$0.01     &$\mp$0.02     &  & 0.25    &$\cdots\;\;$    &$\pm$0.01      &$\pm$0.07       &$\pm$0.01       &$\mp$0.01     \\
$M_2/\msun$        &   48.7   &$\mp$13.6\pa\pa&$\mp$4.5\pa   &$\mp$9.7\pa    &$\mp$4.6\pa   &$\cdots\;\;$  &  & 50.2    &$\mp$14.2\pa\pa &$\mp$6.8\pa    &$\mp$10.2\pa\pa &$\mp$5.4\pa     &$\cdots\;\;$  \\
$\log(L_2/\lsun)$  &   5.10   &$\mp$0.08      &$\cdots\;\;$  & $\pm$0.18     &$\cdots\;\;$  & $\pm$0.06    &  & 5.00    &$\mp$0.08       &$\cdots\;\;$   & $\pm$0.18      &$\cdots\;\;$    & $\pm$0.06    \\
$\logg_2$ (cgs)    &   4.11   &$\mp$0.04      &$\mp$0.04     &$\mp$0.27      &$\mp$0.04     & $\pm$0.04    &  & 4.28    &$\mp$0.04       &$\mp$0.06      &$\mp$0.27       &$\mp$0.05       & $\pm$0.04    \\
\vcssini\ (\kms)    &   52     &$\cdots\;\;$   & $\pm$3       & $\pm$14       & $\pm$2       &$\mp$3        &  & 33      &$\cdots\;\;$    & $\pm$2        & $\pm$9         & $\pm$1         &$\mp$1        \\
\\                                                                                                                                                                                              
$\sin(i)$  \B      &   0.70   & $\pm$0.07     & $\pm$0.04    & $\pm$0.05     & $\pm$0.02    &$\cdots\;\;$  &  & 0.66    & $\pm$0.07      & $\pm$0.04     & $\pm$0.04      & $\pm$0.02      &$\cdots\;\;$  \\
\hline
\end{tabular}
\tablefoot{The first five rows are input parameters, while subsequent
  entries are derived quantities, obtained by assuming that the
  primaries fill their Roche lobes.  The baseline first-estimate
  models (columns headed `M1') result from adopting spectroscopic
  results directly; these models are discussed in
  Section~\ref{s:syspar}.  The `\vcxsini' rows list the projected
  equatorial corotation velocities (obtained by assuming that orbital
  and rotational angular momenta are aligned).  Subsequent columns
  illustrate the sensitivities of derived quantities to moderate
  changes in input parameters.  (The spectroscopic
  analysis may not faithfully reflect the secondary photospheres, in
  which case the parameters are subject to larger uncertainties; 
cf.~$\S$\ref{sec_psp}.)
}
\label{tab:specres2}
\end{sidewaystable*}

\subsection{A first estimate of system parameters from spectroscopy}
\label{s:syspar}

The substantial photometric variability strongly suggests that the
primaries fill, or very nearly fill, their Roche lobes, as 
do the various indicators of lobe-overflow mass transfer,
discussed further below ($\S$\ref{sec_psp}).  With this assumption, and using the procedures
outlined in Section~\ref{sec:syscon}, we can make a first estimate of
approximate actual system parameters, which are summarized in
Table~\ref{tab:specres2} (columns headed `M1').  The Table also
explores the sensitivity of derived quantities to input parameters.
Masses are the least well determined variables, principally because of
the third-power dependence on $\sin{i}$.

The inferred inclinations are consistent with the absence of clear
eclipses, and there is tolerable agreement between the orbital and
model-atmosphere estimates of $\log{g}$, although the spectroscopic
determinations are $\sim$0.3~dex smaller.\footnote{Corrections for
  centrifugal forces, $\Delta\log{g} \simeq (\vesini)^2/R_*$, are
  $\lesssim$0.03.}  For these first-pass parameter estimates,
projected equatorial corotation velocities are in good agreement with
the primaries' observed values, but the secondaries appear to be
rotating considerably faster than synchronous (although well below
critical).

\subsection{Light-curves}

Model light-curves for parameters in the region of the `M1' solutions
fail to reproduce the amplitudes of the observed light-curves.  The
observed `ellipsoidal' variations (Fig.~\ref{fig:photom}) imply that
the primary in each system must be very close to filling its Robe
lobe, but if the secondary is hotter and fainter than the primary,
while being more massive, then it \emph{must} significantly
under\-fill its Roche lobe, regardless of detailed numerical parameter
values.  The amplitude of orbital photo\-metric variability under
these circumstances does not substantially exceed $\sim$0{\fm}1 over
a range of mass ratios and inclinations -- about half the observed
amplitudes.  Varying the spectroscopically inferred parameters over
plausible ranges cannot overcome this discrepancy; the only way to
reproduce the light-curve amplitude by conventional models is to adopt
an overcontact (or double-contact) configuration, but this would imply
spectroscopically more conspicuous secondaries.

\begin{figure}[htpb]
\begin{center}
\includegraphics[angle=270, scale=.39]{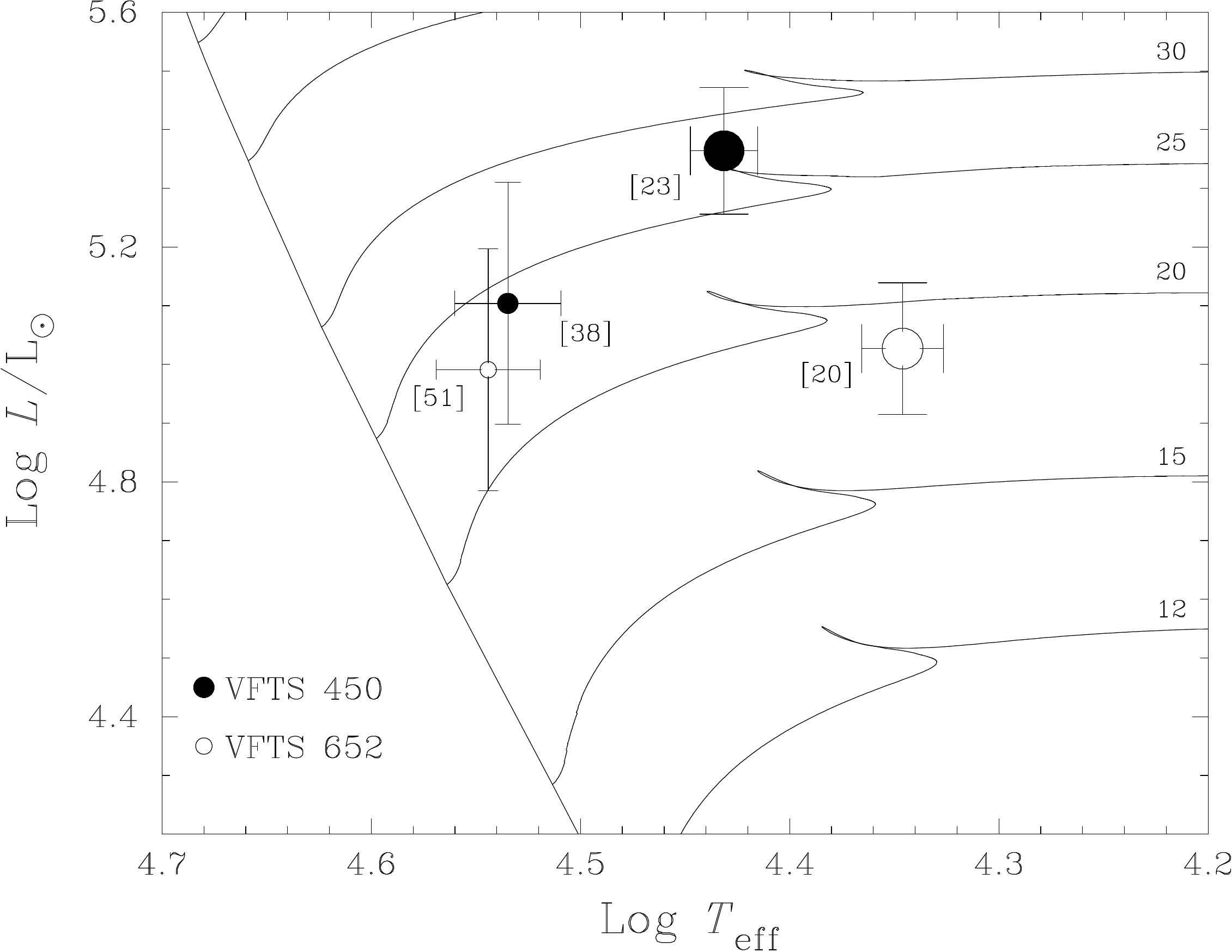} 
\caption{Stellar parameters 
from Table~\ref{tab:specres2}, plotted in the
Hertzprung--Russell diagram.   Error bars illustrate uncertainties 
of $\pm{1}$~kK on primaries
(larger symbols) and $\pm{2}$~kK on secondaries, and the sums in
quadrature of the error ranges on $L$ listed in Table~\ref{tab:specres2}.
Dynamical masses are indicated in square
brackets.
Evolutionary tracks from \citet{bro11a} for single, non-rotating stars
are shown for comparison, labelled by ZAMS mass.}
\label{f_dis_HRD}
\end{center}
\end{figure}

\begin{figure*}
\begin{center}
\includegraphics[scale=0.92]{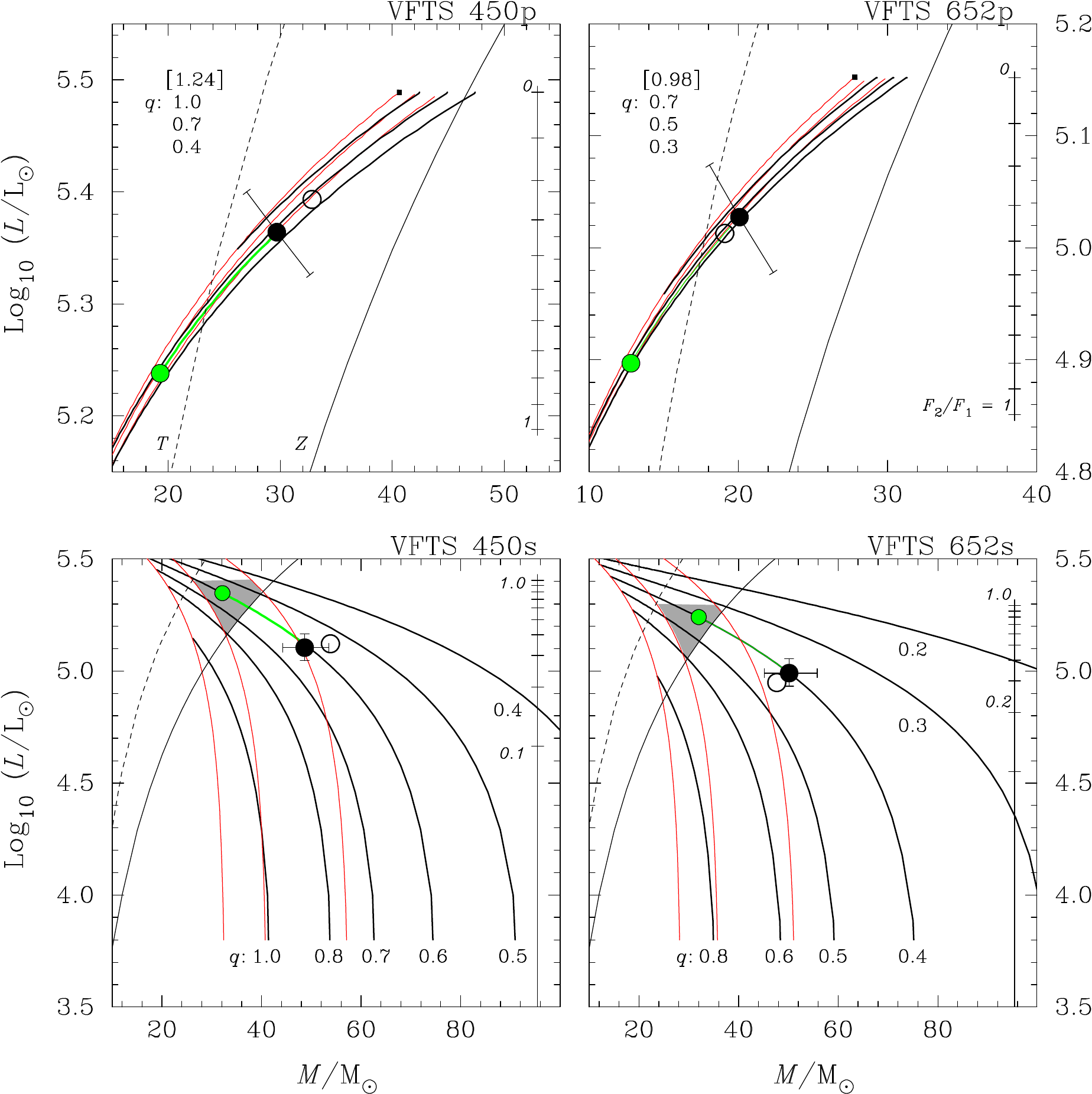}
\caption{Constraints in the mass--luminosity plane for the primary and
  secondary components (upper, lower panels) of VFTS~450 and VFTS~652
  (left, right panels), obtained by assuming that the primaries fill their Roche
  lobes, with absolute magnitudes and effective temperatures as
summarized in Table~\ref{tab:specres2};
refer to Section~\ref{sec:con1} for
  further details.
\newline Thick black solid lines
  show $M$--$L$ loci for the indicated mass ratios $q$ over a
  range in continuum brightness ratio, \fstwo/\fsone\ (values shown in
  the vertical scales to the right in each panel, marked at steps of
  0.1; $\log({L}/\lsun)$ is constant for given \fstwo/\fsone, for
  fixed \teff\ values).  
Thin solid and dashed curves, labelled $Z$
  and $T$ in the top-left panel, show the zero-age and terminal-age
  main sequence loci for non-rotating single stars (from
  \citealt{bro11a,bro11b}).  Thin red curves are lines of constant
  inclination, at $i = 90$, 60, and~45$^\circ$ (left to right).  Grey
  shaded areas in the lower panels indicate the zones for which
  $45^\circ \le i \le 60^\circ$, $L \ge L\mbox{(ZAMS)}$, and
  $\fstwo/\fsone \le 1$.\newline Filled circles show the initial
  parameter estimates summarised in Table~\ref{tab:specres2} (columns
  headed `M1'). `Error bars' in the upper panels, and horizontal error 
  bars in the lower panels, show the effects of changing \teff(1) by
  $\pm$1~kK (this affects the inferred secondary mass, but not its
  luminosity, all else fixed).  Vertical error bars in the lower
  panels show the effect of varying \teff(2) by $\pm$2~kK (which has
  no effect on secondary mass).\newline 
 Open circles represent equivalent $M$--$L$ solutions from $V$-band
  photometry. Green circles show the effects of (arbitrarily)
  adjusting \fstwo/\fsone\ to bring the secondary masses to 32~\msun,
  hence into the grey shaded zones in this plane.\newline
[Note that any changes to
    effective temperatures or absolute magnitudes 
also change the loci of constant $q$, so that {\emph {only}}
$M$ and $L$ can be inferred from this diagram 
for \teff\ or $M(I_{\rm C})$ values that differ from the reference
    solution;  e.g., the $V$-band solutions have the same $q$,
    $i$ values as the $I_{\rm C}$-band solutions.]}
\label{f_masslum}
\end{center}
\end{figure*}

\subsection{Evolutionary considerations}
\label{s_dis_evo}

The schematic `M1' system parameters
are plotted in an \mbox{H--R} diagram in
Fig.~\ref{f_dis_HRD}; evolutionary tracks for non-rotating single
stars at LMC metallicity, from \citet{bro11a,bro11b}, are also shown.
This figure discloses a further problem:
although the dynamical masses estimated for the primaries (i.e., the cooler,
less massive, lobe-filling components) are in reasonably good agreement with the single-star
tracks, the secondaries are significantly under-luminous for their
dynamical masses -- and standard binary evolution cannot produce this
outcome.

Nevertheless, it is unlikely that these systems can be anything other
than hot, massive counterparts of typical Algol-type binaries, in the
slow (nuclear-timescale) phase of Case~A mass transfer.  Mass transfer
(including common-envelope evolution) in a more evolved configuration
would produce a helium star (a Wolf-Rayet star at these masses), which
would be spectroscopically conspicuous. 
Chemically homogeneous
evolution of the rapidly rotating secondaries would lead to
significantly higher effective temperatures, and can probably also be 
excluded.

\begin{figure*}
\begin{center}
\includegraphics[scale=0.92,angle=270]{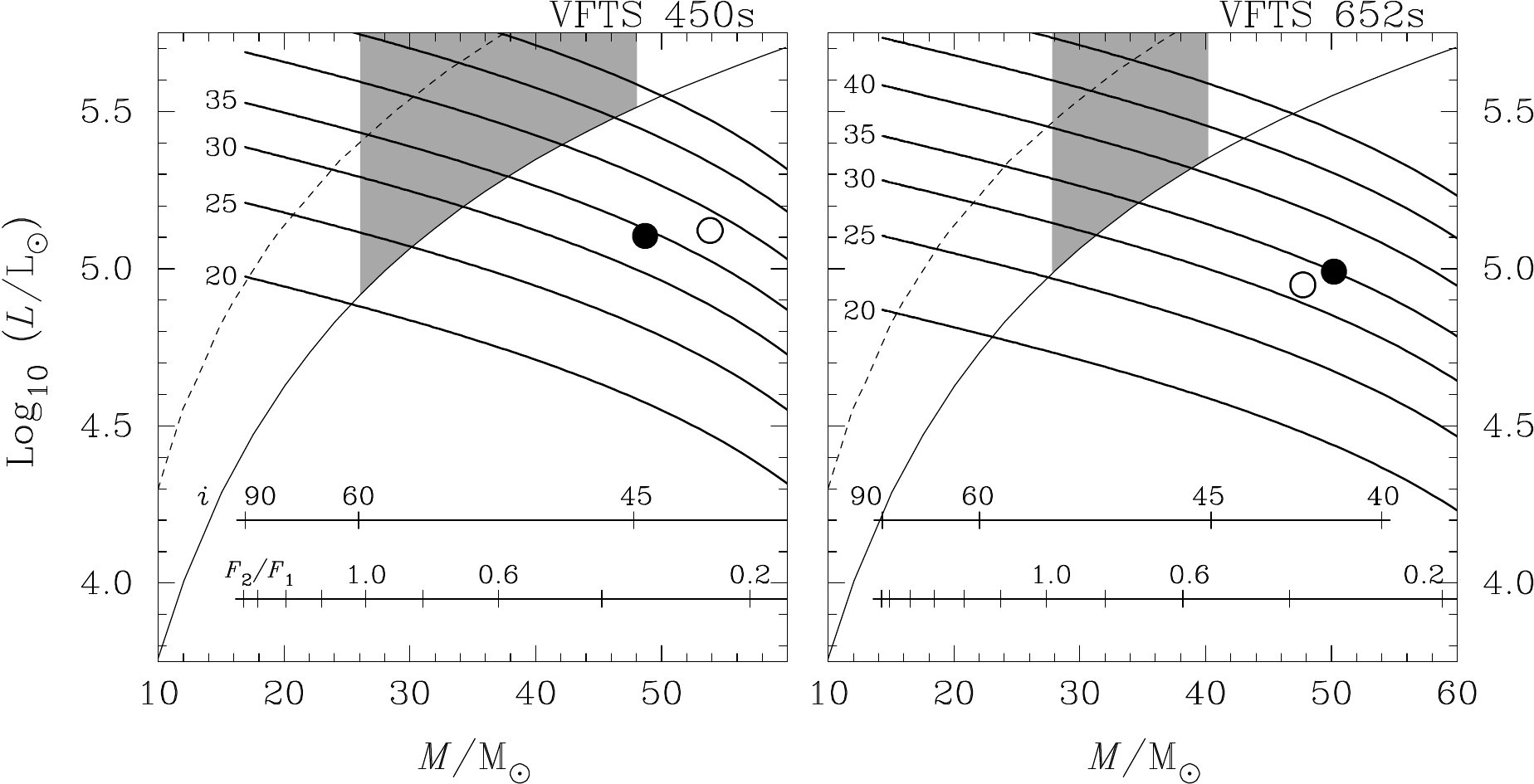}
\caption{Constraints in the mass--luminosity plane for secondary
  components for fixed $q$, \teff(1), and absolute magnitudes, obtained
  by assuming that the primaries fill their Roche lobes
  (cp.\ Fig.~\ref{f_masslum});
refer to Section~\ref{sec:con2} for details.
\newline Thick black solid lines show $M$--$L$ loci for the indicated
secondary temperatures (in~kK) over a range in continuum brightness
ratio, \fstwo/\fsone\ (values shown in the lower horizontal scales,
marked at steps of 0.2).  The secondary mass $M$ is constant for given
\fstwo/\fsone (for fixed $q$ values), as is the orbital inclination
(upper horizontal scales, labelled in degrees).  Thin solid and dashed
curves show the zero-age
and terminal-age main sequence loci for non-rotating single stars
(from \citealt{bro11a,bro11b}).   Grey
shaded areas indicate the zones for which
$45^\circ \le i \le 60^\circ$, $L \ge L\mbox{(ZAMS)}$, and
$\fstwo/\fsone \le 1$.
\newline Filled circles show the initial
  parameter estimates summarised in Table~\ref{tab:specres2} (columns
  headed `M1'). 
 Open circles represent equivalent results for $V$-band
  photometry.}
\label{f_masslum2}
\end{center}
\end{figure*}

\subsection{Parameter-space exploration}
\label{sec_psp}

Given the difficulties encountered in reconciling spectroscopic,
photometric, and evolutionary constraints, the finger of doubt points
most directly at the unqualified attribution of the observed secondary
spectra to the secondary stars' photo\-spheres.  Taken together, the
breadth of the absorption profiles ($\S$\ref{sec:vsini}), the
discrepancies in radial velocities from different lines
($\S$\ref{sec:sspec}), the anomalous double-peaked H$\alpha$ profiles
(Fig.~\ref{f_redspec}), unusual near-IR colours
($\S$\ref{s_obs_photo2}), and the general elusiveness of the secondary
spectra, all suggest the possibility that the secondary's photospheric
spectrum in each system may be modified, or even concealed, by an accretion
disk (which would also be consistent with lobe-filling primaries).

These observed properties are reminiscent of the W~Ser class of
binaries \citep{plavec80, tarasov00}.  Although the VFTS targets have
higher masses and shorter periods than is typical for this group,
their general characteristics, including lobe-filling, synchronously
rotating primaries, and IR excesses, are in accord with this notion
\citep{anders89,menni10}, and there are clear similarities to related
early-type systems such as RY~Sct and V453~Sco
\citep{grund07,josephs01}.

The VFTS binaries
studied here may well, therefore, have secondary
spectra that are contaminated by, or arise in, accretion disks.
In recognition of this possibility, we explore a broader parameter
space for the systems, and particularly for the secondary components.

\subsubsection{Mass ratios, brightness ratios}
\label{sec:con1}

For heuristic purposes,
we first consider the consequences of adopting mass ratios $q$ and
continuum brightness ratios \fstwo/\fsone\ in ranges outside those
directly inferred from the secondary spectra.  
As described in
Section~\ref{sec:syscon}, a brightness ratio and component
temperatures yield the stellar radii; given the primary radius, a mass
ratio gives the inclination (for a lobe-filling primary), and hence,
from the spectroscopic orbit, the masses.  Luminosities follow from
the radii and temperatures.

The basic inputs we adopt for each system are therefore (i) the
absolute magnitude; (ii) the orbital period; (iii) the primary's
orbital-velocity semi-amplitude; and (iv) the two components' effective
temperatures.  With fixed values for these quantities, the system
characteristics are fully specified by $q$ and
\fstwo/\fsone\ (assuming that the primary fills its Roche lobe).

Results in the mass--luminosity plane are illustrated in
Fig.~\ref{f_masslum}.  In this figure, $M$--$L$ curves are shown for
selected specific values of the mass ratio $q$, over a range in
\fstwo/\fsone.  The curves are insensitive to $q$ for the primary
stars, but not for the secondaries.

Also shown in the Figure are lines of constant orbital
inclination. For a given $q$, then a particular inclination
corresponds to a specific \fstwo/\fsone; this \{$q$, \fstwo/\fsone\}
pair yields the full set of other parameters, including $M$ and $L$.
Thus a constant inclination corresponds to a curve in the $M$--$L$
plane (again, assuming that the primary fills its Roche lobe).

Other than at advanced evolutionary stages, a single hot star will
normally lie between the ZAMS and TAMS $M$--$L$ loci in
Fig.~\ref{f_masslum}.\footnote{This is true even shortly after leaving
  the main sequence, as an isolated massive star evolves to the right
  in the Hertzsprung--Russell diagram at almost constant mass and
  luminosity (cf., e.g., Fig.~\ref{f_dis_HRD}).}  If binary evolution
is to produce a secondary that is not underluminous for its mass, then this
component must lie somewhere above the ZAMS $M$--$L$ locus in
Fig.~\ref{f_masslum}.  Moreover, the absence of obvious eclipses,
coupled with significant ellipsoidal-type photometric variability,
suggests $60^\circ \gtrsim i \gtrsim 45^\circ$.  Finally, it seems
reasonable to suppose $\fstwo/\fsone < 1$. The areas marked in grey in
Fig.~\ref{f_masslum} meet these three constraints.

For each system, solutions that move the secondary into this grey zone
can be achieved by increasing \fstwo/\fsone; by increasing both the
mass ratio and $\teff$(2) over the default M1
values; or some combination of these.

Implausibly large increases in secondary temperature are required to
make the secondaries sufficiently luminous.  Tolerable solutions
are, however, possible by
adopting values of \fstwo/\fsone\ that are somewhat larger than those
inferred from the spectroscopy; this is not unreasonable given the
errors, and the possibility that the secondary absorption-line spectra
are `veiled' by circumstellar material.

Although an increase in $\fstwo/\fsone$ suggests a compensating
decrease in the secondary effective temperature
(Table~\ref{t_comparison}), we don't attempt to refine the details
further, given the considerable uncertainties (and noting that
reductions of only $\lesssim 10$\%\ in \teff\ are permitted if, at $i
\lesssim 60^\circ$, the secondaries are not to be underluminous for
their masses), but merely conclude that consistency between
observations and broad evolutionary considerations can be achieved by
plausible adjustments to the flux ratios, while retaining values for
other parameters that are close to those estimated spectroscopically.


\subsubsection{Secondary temperatures, brightness ratios}
\label{sec:con2}

Secondary temperatures are particularly prone to uncertainty if the
secondary spectra are not purely photospheric, so we also explicitly examine the
consequences of treating \teff(2) as a variable.  For each system, the
basic inputs are again (i) the absolute magnitude; (ii)
the orbital period; (iii) the primary's orbital-velocity
semi-amplitude; with additionally (iv) the mass ratio and (v) the primary
temperature fixed at selected values.  

System characteristics are now fully specified by \teff(2) and
\fstwo/\fsone\ (assuming that the primary fills its Roche lobe).
The consequences of varying these parameters are illustrated in
Fig.~\ref{f_masslum2}.  Once again, the simplest way to migrate the
secondaries into the `zone of plausibility' (shown in grey in the
Figure) is to increase \fstwo/\fsone, and/or to increase \teff(2).  We
conclude that, most probably, the secondary components are brighter
than is superficially suggested by the spectra, but that their spectra
are veiled by accretion disks.  The likelihood is that they are then
also larger, and closer to filling their Roche lobes, than suggested by
the M1 parameters, which could
reconcile system properties with the photometry.

\section{Summary}

We have presented new spectroscopy of the massive blue binaries
VFTS~450 and VFTS~652.  Well-determined orbits are established for the
spectroscopically more conspicuous components in both systems (the
`primaries' in our notation; Table~\ref{tab:specorb}); we argue
that these are the less massive
components, and that they fill their Roche lobes, with
near-synchronous rotation.  Model-atmosphere
analyses of the primaries yield reasonably robust results
(Table~\ref{t_comparison}), demonstrating significant
surface-nitrogen abundances in each case.

The secondary spectra have been detected, although the inferred
characteristics are considerably less well established.  If these
secondary spectra reliably reflect the photospheric properties of the secondary
stars, then they are associated with the hotter
components.   However,
quantitative models of the systems built on the spectroscopic results
(Table~\ref{tab:specres2}) have
inconsistencies with the photo\-metry, and with evolutionary considerations, as
discussed in Sections~\ref{s_dis_evo} and~\ref{sec_psp}.  
We suggest that the secondary spectra 
are contaminated by, or arise in, accretion disks, and have explored
the consequences of relaxing the allowed values for relevant
`observed' secondary parameters (Figs.~\ref{f_masslum}, \ref{f_masslum2}).

\begin{acknowledgements}

We thank our anonymous referee for a careful reading of the paper, and
helpful suggestions.  LAA acknowledges support from the
Funda\c{c}\~{a}o de Amparo \`a Pesquisa do Estado de S\~{a}o Paulo 
(FAPESP, N2013/18245-0 and \mbox{2012/09716-6}).  AZB acknowledges research
and travel support from the European Commission Framework Program
Seven under Marie Curie International Reintegration Grant
PIRG04-GA-2008-239335.  SS-D acknowledges financial support from the
Spanish Ministry of Economy and Competitiveness (MINECO) under grants
AYA2010-21697-C05-04, and Severo Ochoa SEV-2011-0187, and by the
Canary Islands Government under grant PID2010119.  OGLE work on binary
systems is partly supported by Polish National Science Centre grant
no. DEC-2011/03/B/ST9/02573.

\end{acknowledgements}

%
\bibliographystyle{aa}

\bibliography{Bbinary}

\begin{thebibliography}{49}
\expandafter\ifx\csname natexlab\endcsname\relax\def\natexlab#1{#1}\fi

\bibitem[{{Andersen} \& {Nordstr\"om}(1989)}]{anders89}
{Andersen}, J. \& {Nordstr\"om}, B. 1989, \ssr, 50, 179

\bibitem[{{Bonanos}(2009)}]{bon09a}
{Bonanos}, A.~Z. 2009, \apj, 691, 407

\bibitem[{{Bonanos} {et~al.}(2009){Bonanos}, {Massa}, {Sewilo}, {Lennon},
  {Panagia}, {Smith}, {Meixner}, {Babler}, {Bracker}, {Meade}, {Gordon},
  {Hora}, {Indebetouw}, \& {Whitney}}]{bon09b}
{Bonanos}, A.~Z., {Massa}, D.~L., {Sewilo}, M., {et~al.} 2009, \aj, 138, 1003

\bibitem[{{Brott} {et~al.}(2011{\natexlab{a}}){Brott}, {de Mink}, {Cantiello},
  {Langer}, {de Koter}, {Evans}, {Hunter}, {Trundle}, \& {Vink}}]{bro11a}
{Brott}, I., {de Mink}, S.~E., {Cantiello}, M., {et~al.} 2011{\natexlab{a}},
  \aap, 530, A115

\bibitem[{{Brott} {et~al.}(2011{\natexlab{b}}){Brott}, {Evans}, {Hunter}, {de
  Koter}, {Langer}, {Dufton}, {Cantiello}, {Trundle}, {Lennon}, {de Mink},
  {Yoon}, \& {Anders}}]{bro11b}
{Brott}, I., {Evans}, C.~J., {Hunter}, I., {et~al.} 2011{\natexlab{b}}, \aap,
  530, A116

\bibitem[{{Dufton} {et~al.}(2013){Dufton}, {Langer}, {Dunstall}, {Evans},
  {Brott}, {de Mink}, {Howarth}, {Kennedy}, {McEvoy}, {Potter},
  {Ram{\'{\i}}rez-Agudelo}, {Sana}, {Sim{\'o}n-D{\'{\i}}az}, {Taylor}, \&
  {Vink}}]{duf12}
{Dufton}, P.~L., {Langer}, N., {Dunstall}, P.~R., {et~al.} 2013, \aap, 550,
  A109

\bibitem[{{Dufton} {et~al.}(2005){Dufton}, {Ryans}, {Trundle}, {Lennon},
  {Hubeny}, {Lanz}, \& {Allende Prieto}}]{duf05}
{Dufton}, P.~L., {Ryans}, R.~S.~I., {Trundle}, C., {et~al.} 2005, \aap, 434,
  1125

\bibitem[{{Dunstall} {et~al.}(2015){Dunstall}, {Dufton}, {Sana}, {Evans},
  {Howarth}, {Sim{\'o}n-D{\'{\i}}az}, {de Mink}, {Langer}, {Ma{\'{\i}}z
  Apell{\'a}niz}, \& {Taylor}}]{dunstall15}
{Dunstall}, P.~R., {Dufton}, P.~L., {Sana}, H., {et~al.} 2015, ArXiv e-prints

\bibitem[{{Eggleton}(1983)}]{Eggleton83}
{Eggleton}, P.~P. 1983, \apj, 268, 368

\bibitem[{{Evans} {et~al.}(2011){Evans}, {Taylor}, {H{\'e}nault-Brunet},
  {Sana}, {de Koter}, {Sim{\'o}n-D{\'{\i}}az}, {Carraro}, {Bagnoli}, {Bastian},
  {Bestenlehner}, {Bonanos}, {Bressert}, {Brott}, {Campbell}, {Cantiello},
  {Clark}, {Costa}, {Crowther}, {de Mink}, {Doran}, {Dufton}, {Dunstall},
  {Friedrich}, {Garcia}, {Gieles}, {Gr{\"a}fener}, {Herrero}, {Howarth},
  {Izzard}, {Langer}, {Lennon}, {Ma{\'{\i}}z Apell{\'a}niz}, {Markova},
  {Najarro}, {Puls}, {Ramirez}, {Sab{\'{\i}}n-Sanjuli{\'a}n}, {Smartt},
  {Stroud}, {van Loon}, {Vink}, \& {Walborn}}]{eva11}
{Evans}, C.~J., {Taylor}, W.~D., {H{\'e}nault-Brunet}, V., {et~al.} 2011, \aap,
  530, A108

\bibitem[{{Ferraz-Mello}(1981)}]{ferraz81}
{Ferraz-Mello}, S. 1981, \aj, 86, 619

\bibitem[{{Grundstrom} {et~al.}(2007){Grundstrom}, {Gies}, {Hillwig},
  {McSwain}, {Smith}, {Gehrz}, {Stahl}, \& {Kaufer}}]{grund07}
{Grundstrom}, E.~D., {Gies}, D.~R., {Hillwig}, T.~C., {et~al.} 2007, \apj, 667,
  505

\bibitem[{{Hadrava}(2004)}]{hadrava04}
{Hadrava}, P. 2004, in Astronomical Society of the Pacific Conference Series,
  Vol. 318, Spectroscopically and Spatially Resolving the Components of the
  Close Binary Stars, ed. R.~W. {Hilditch}, H.~{Hensberge}, \& K.~{Pavlovski},
  86--94

\bibitem[{{Howarth}(1983)}]{howarth83}
{Howarth}, I.~D. 1983, \mnras, 203, 301

\bibitem[{{Howarth}(2011)}]{Howarth11a}
{Howarth}, I.~D. 2011, \mnras, 413, 1515

\bibitem[{{Hubeny}(1988)}]{hub88}
{Hubeny}, I. 1988, Computer Physics Communications, 52, 103

\bibitem[{{Hubeny} {et~al.}(1998){Hubeny}, {Heap}, \& {Lanz}}]{hub98}
{Hubeny}, I., {Heap}, S.~R., \& {Lanz}, T. 1998, in Astronomical Society of the
  Pacific Conference Series, Vol. 131, Properties of Hot Luminous Stars, ed.
  {I.~Howarth}, 108

\bibitem[{{Hubeny} \& {Lanz}(1995)}]{hub95}
{Hubeny}, I. \& {Lanz}, T. 1995, \apj, 439, 875

\bibitem[{{Hunter} {et~al.}(2007){Hunter}, {Dufton}, {Smartt}, {Ryans},
  {Evans}, {Lennon}, {Trundle}, {Hubeny}, \& {Lanz}}]{hun07}
{Hunter}, I., {Dufton}, P.~L., {Smartt}, S.~J., {et~al.} 2007, \aap, 466, 277

\bibitem[{{Iliji\'c}(2004)}]{Ilijic04}
{Iliji\'c}, S. 2004, in Astronomical Society of the Pacific Conference Series,
  Vol. 318, Spectroscopically and Spatially Resolving the Components of the
  Close Binary Stars, ed. R.~W. {Hilditch}, H.~{Hensberge}, \& K.~{Pavlovski},
  107--110

\bibitem[{{Josephs} {et~al.}(2001){Josephs}, {Gies}, {Bagnuolo}, {Shure},
  {Penny}, \& {Wang}}]{josephs01}
{Josephs}, T.~S., {Gies}, D.~R., {Bagnuolo}, Jr., W.~G., {et~al.} 2001, \pasp,
  113, 957

\bibitem[{{Kato} {et~al.}(2007){Kato}, {Nagashima}, {Nagayama}, {Kurita},
  {Koerwer}, {Kawai}, {Yamamuro}, {Zenno}, {Nishiyama}, {Baba}, {Kadowaki},
  {Haba}, {Hatano}, {Shimizu}, {Nishimura}, {Nagata}, {Sato}, {Murai},
  {Kawazu}, {Nakajima}, {Nakaya}, {Kandori}, {Kusakabe}, {Ishihara},
  {Kaneyasu}, {Hashimoto}, {Tamura}, {Tanab{\'e}}, {Ita}, {Matsunaga},
  {Nakada}, {Sugitani}, {Wakamatsu}, {Glass}, {Feast}, {Menzies}, {Whitelock},
  {Fourie}, {Stoffels}, {Evans}, \& {Hasegawa}}]{kat07}
{Kato}, D., {Nagashima}, C., {Nagayama}, T., {et~al.} 2007, \pasj, 59, 615

\bibitem[{{Lucy} \& {Sweeney}(1971)}]{Lucy71}
{Lucy}, L.~B. \& {Sweeney}, M.~A. 1971, \aj, 76, 544

\bibitem[{{McEvoy} {et~al.}(2015){McEvoy}, {Dufton}, {Evans}, {Kalari},
  {Markova}, {Sim{\'o}n-D{\'{\i}}az}, {Vink}, {Walborn}, {Crowther}, {de
  Koter}, {de Mink}, {Dunstall}, {H{\'e}nault-Brunet}, {Herrero}, {Langer},
  {Lennon}, {Ma{\'{\i}}z Apell{\'a}niz}, {Najarro}, {Puls}, {Sana},
  {Schneider}, \& {Taylor}}]{mce15}
{McEvoy}, C.~M., {Dufton}, P.~L., {Evans}, C.~J., {et~al.} 2015, \aap, 575, A70

\bibitem[{{Meixner} {et~al.}(2006){Meixner}, {Gordon}, {Indebetouw}, {Hora},
  {Whitney}, {Blum}, {Reach}, {Bernard}, {Meade}, {Babler}, {Engelbracht},
  {For}, {Misselt}, {Vijh}, {Leitherer}, {Cohen}, {Churchwell}, {Boulanger},
  {Frogel}, {Fukui}, {Gallagher}, {Gorjian}, {Harris}, {Kelly}, {Kawamura},
  {Kim}, {Latter}, {Madden}, {Markwick-Kemper}, {Mizuno}, {Mizuno}, {Mould},
  {Nota}, {Oey}, {Olsen}, {Onishi}, {Paladini}, {Panagia}, {Perez-Gonzalez},
  {Shibai}, {Sato}, {Smith}, {Staveley-Smith}, {Tielens}, {Ueta}, {van Dyk},
  {Volk}, {Werner}, \& {Zaritsky}}]{mei06}
{Meixner}, M., {Gordon}, K.~D., {Indebetouw}, R., {et~al.} 2006, \aj, 132, 2268

\bibitem[{{Melnick}(1985)}]{mel85}
{Melnick}, J. 1985, A\&A, 153, 235

\bibitem[{{Mennickent} \& {Ko{\l}aczkowski}(2010)}]{menni10}
{Mennickent}, R.~E. \& {Ko{\l}aczkowski}, Z. 2010, in Revista Mexicana de
  Astronomia y Astrofisica, vol. 27, Vol.~38, Revista Mexicana de Astronomia y
  Astrofisica Conference Series, 23--26

\bibitem[{{Niemela} \& {Bassino}(1994)}]{niemala94}
{Niemela}, V.~S. \& {Bassino}, L.~P. 1994, \apj, 437, 332

\bibitem[{{Niemela} \& {Morrell}(1986)}]{niemela86}
{Niemela}, V.~S. \& {Morrell}, N.~I. 1986, \apj, 310, 715

\bibitem[{{Pasquini} {et~al.}(2002){Pasquini}, {Avila}, {Blecha}, {Cacciari},
  {Cayatte}, {Colless}, {Damiani}, {de Propris}, {Dekker}, {di Marcantonio},
  {Farrell}, {Gillingham}, {Guinouard}, {Hammer}, {Kaufer}, {Hill}, {Marteaud},
  {Modigliani}, {Mulas}, {North}, {Popovic}, {Rossetti}, {Royer}, {Santin},
  {Schmutzer}, {Simond}, {Vola}, {Waller}, \& {Zoccali}}]{pas02}
{Pasquini}, L., {Avila}, G., {Blecha}, A., {et~al.} 2002, The Messenger, 110, 1

\bibitem[{{Pietrzy{\'n}ski} {et~al.}(2013){Pietrzy{\'n}ski}, {Graczyk},
  {Gieren}, {Thompson}, {Pilecki}, {Udalski}, {Soszy{\'n}ski}, {Koz{\l}owski},
  {Konorski}, {Suchomska}, {Bono}, {Moroni}, {Villanova}, {Nardetto},
  {Bresolin}, {Kudritzki}, {Storm}, {Gallenne}, {Smolec}, {Minniti}, {Kubiak},
  {Szyma{\'n}ski}, {Poleski}, {Wyrzykowski}, {Ulaczyk}, {Pietrukowicz},
  {G{\'o}rski}, \& {Karczmarek}}]{Pietr13}
{Pietrzy{\'n}ski}, G., {Graczyk}, D., {Gieren}, W., {et~al.} 2013, \nat, 495,
  76

\bibitem[{{Plavec}(1980)}]{plavec80}
{Plavec}, M.~J. 1980, in IAU Symposium, Vol.~88, Close Binary Stars:
  Observations and Interpretation, ed. M.~J. {Plavec}, D.~M. {Popper}, \& R.~K.
  {Ulrich}, 251--260

\bibitem[{{Ram{\'{\i}}rez-Agudelo} {et~al.}(2013){Ram{\'{\i}}rez-Agudelo},
  {Sim{\'o}n-D{\'{\i}}az}, {Sana}, {de Koter}, {Sab{\'{\i}}n-Sanjul{\'{\i}}an},
  {de Mink}, {Dufton}, {Gr{\"a}fener}, {Evans}, {Herrero}, {Langer}, {Lennon},
  {Ma{\'{\i}}z Apell{\'a}niz}, {Markova}, {Najarro}, {Puls}, {Taylor}, \&
  {Vink}}]{ram13}
{Ram{\'{\i}}rez-Agudelo}, O.~H., {Sim{\'o}n-D{\'{\i}}az}, S., {Sana}, H.,
  {et~al.} 2013, \aap, 560, A29

\bibitem[{{Ryans} {et~al.}(2003){Ryans}, {Dufton}, {Mooney}, {Rolleston},
  {Keenan}, {Hubeny}, \& {Lanz}}]{rya03}
{Ryans}, R.~S.~I., {Dufton}, P.~L., {Mooney}, C.~J., {et~al.} 2003, \aap, 401,
  1119

\bibitem[{{Sana} {et~al.}(2013){Sana}, {de Koter}, {de Mink}, {Dunstall},
  {Evans}, {H{\'e}nault-Brunet}, {Ma{\'{\i}}z Apell{\'a}niz},
  {Ram{\'{\i}}rez-Agudelo}, {Taylor}, {Walborn}, {Clark}, {Crowther},
  {Herrero}, {Gieles}, {Langer}, {Lennon}, \& {Vink}}]{sana13}
{Sana}, H., {de Koter}, A., {de Mink}, S.~E., {et~al.} 2013, \aap, 550, A107

\bibitem[{{Schaefer}(2008)}]{Schaefer08}
{Schaefer}, B.~E. 2008, \aj, 135, 112

\bibitem[{{Schechter} {et~al.}(1993){Schechter}, {Mateo}, \& {Saha}}]{dophot}
{Schechter}, P.~L., {Mateo}, M., \& {Saha}, A. 1993, \pasp, 105, 1342

\bibitem[{{Selman} {et~al.}(1999){Selman}, {Melnick}, {Bosch}, \&
  {Terlevich}}]{sel99}
{Selman}, F., {Melnick}, J., {Bosch}, G., \& {Terlevich}, R. 1999, \aap, 347,
  532

\bibitem[{{Sim{\'o}n-D{\'{\i}}az} \& {Herrero}(2014)}]{sim14}
{Sim{\'o}n-D{\'{\i}}az}, S. \& {Herrero}, A. 2014, \aap, 562, A135

\bibitem[{{Sota} {et~al.}(2011){Sota}, {Ma{\'{\i}}z Apell{\'a}niz}, {Walborn},
  {Alfaro}, {Barb{\'a}}, {Morrell}, {Gamen}, \& {Arias}}]{sot11}
{Sota}, A., {Ma{\'{\i}}z Apell{\'a}niz}, J., {Walborn}, N.~R., {et~al.} 2011,
  \apjs, 193, 24

\bibitem[{{Sundqvist} {et~al.}(2013){Sundqvist}, {Sim{\'o}n-D{\'{\i}}az},
  {Puls}, \& {Markova}}]{sun13}
{Sundqvist}, J.~O., {Sim{\'o}n-D{\'{\i}}az}, S., {Puls}, J., \& {Markova}, N.
  2013, \aap, 559, L10

\bibitem[{{Tarasov}(2000)}]{tarasov00}
{Tarasov}, A.~E. 2000, in Astronomical Society of the Pacific Conference
  Series, Vol. 214, IAU Colloq. 175: The Be Phenomenon in Early-Type Stars, ed.
  M.~A. {Smith}, H.~F. {Henrichs}, \& J.~{Fabregat}, 644

\bibitem[{{Taylor} {et~al.}(2011){Taylor}, {Evans}, {Sana}, {Walborn}, {de
  Mink}, {Stroud}, {Alvarez-Candal}, {Barb{\'a}}, {Bestenlehner}, {Bonanos},
  {Brott}, {Crowther}, {de Koter}, {Friedrich}, {Gr{\"a}fener},
  {H{\'e}nault-Brunet}, {Herrero}, {Kaper}, {Langer}, {Lennon}, {Ma{\'{\i}}z
  Apell{\'a}niz}, {Markova}, {Morrell}, {Monaco}, \& {Vink}}]{tay11}
{Taylor}, W.~D., {Evans}, C.~J., {Sana}, H., {et~al.} 2011, \aap, 530, L10

\bibitem[{{Udalski} {et~al.}(2008){Udalski}, {Szymanski}, {Soszynski}, \&
  {Poleski}}]{uda08}
{Udalski}, A., {Szymanski}, M.~K., {Soszynski}, I., \& {Poleski}, R. 2008,
  \actaa, 58, 69

\bibitem[{{Udalski} {et~al.}(2015){Udalski}, {Szyma{\'n}ski}, \&
  {Szyma{\'n}ski}}]{uda15}
{Udalski}, A., {Szyma{\'n}ski}, M.~K., \& {Szyma{\'n}ski}, G. 2015, \actaa, 65,
  1

\bibitem[{{Walborn} \& {Blades}(1997)}]{wal97}
{Walborn}, N.~R. \& {Blades}, J.~C. 1997, \apjs, 112, 457

\bibitem[{{Walborn} {et~al.}(2014){Walborn}, {Sana}, {Sim{\'o}n-D{\'{\i}}az},
  {Ma{\'{\i}}z Apell{\'a}niz}, {Taylor}, {Evans}, {Markova}, {Lennon}, \& {de
  Koter}}]{wal14}
{Walborn}, N.~R., {Sana}, H., {Sim{\'o}n-D{\'{\i}}az}, S., {et~al.} 2014, \aap,
  564, A40

\bibitem[{{Wegner}(1994)}]{wegner94}
{Wegner}, W. 1994, \mnras, 270, 229

\bibitem[{{Wilson}(1958)}]{Wilson58}
{Wilson}, R. 1958, Publications of the Royal Observatory of Edinburgh, 2, 62

\end{thebibliography}

\newpage
\setcounter{table}{1}

%
%
%

\footnotesize
\onecolumn
\begin{center}
\begin{longtable}{lcccccccclr}
\caption{\label{t_full_obs1} Log of spectroscopic observations of
    VFTS~450.  Orbital phases are with respect to the circular-orbit
    ephemeris in Table~\ref{tab:specorb}.  The velocity measurements
    are described in Section~\ref{s_rvm}.}\\
\hline
\hline
\multicolumn{1}{c}{\rule{0pt}{8pt}Setting} & MJD \T& Orbital &
\multicolumn{5}{c}{Primary radial velocities (\kms)} &
\multicolumn{2}{c}{Secondary} \\
 & & Phase
& \ion{He}{i} & \ion{He}{i} & \ion{Si}{iii} & \multicolumn{2}{c}{     } &
\multicolumn{2}{c}{RV (\kms)} \\
 & &
& $\lambda$4387 & $\lambda$4471 & $\lambda$4552 & Mean & s.e. &
\multicolumn{2}{c}{\ion{He}{ii} $\lambda$4200} \\
\hline
\endfirsthead
\caption[]{\textit{continued}}\\
\hline
\hline
\multicolumn{1}{c}{\rule{0pt}{8pt}Setting} & MJD \T& Orbital &
\multicolumn{5}{c}{Primary radial velocities (\kms)} &
\multicolumn{2}{c}{Secondary} \\
 & & Phase
& \ion{He}{i} & \ion{He}{i} & \ion{Si}{iii} & \multicolumn{2}{c}{     } &
\multicolumn{2}{c}{RV} \\
 & &
& $\lambda$4387 & $\lambda$4471 & $\lambda$4552 & Mean & s.e. &
\multicolumn{2}{c}{(\kms)} \\
\hline
\endhead
\hline
\endfoot
\hline
\hline
\endlastfoot
\rule{0pt}{10pt}UVES      
          &  54761.2275   &   0.99 & 466.9    &   460.7  &  468.6    &  465.4  &   {\pa}2.4   \\
UVES      &  54761.2491   &   1.00 & 483.1    &   466.0  &  474.6    &  474.6  &   {\pa}5.0   \\
UVES      &  54761.2772   &   0.00 & 468.2    &   476.6  &  455.1    &  466.7  &   {\pa}6.3   \\
UVES      &  54761.2995   &   0.00 & 469.6    &   460.7  &  457.7    &  462.7  &   {\pa}3.6   \\
UVES      &  54767.2738   &   0.87 & 401.9    &   385.0  &  413.6    &  400.1  &   {\pa}8.3   \\
UVES      &  54767.2954   &   0.87 & 381.6    &   383.6  &  399.3    &  388.2  &   {\pa}5.6   \\
UVES      &  54845.1671   &   0.17 & 347.7    &   355.7  &  324.0    &  342.5  &   {\pa}9.5   \\
UVES      &  54845.1888   &   0.18 & 349.1    &   358.4  &  319.8    &  342.4  &       11.6   \\
UVES      &  54876.1214   &   0.66 & 131.1    &   123.3  &  117.6    &  124.0  &   {\pa}3.9   \\
UVES      &  54876.1430   &   0.67 & 128.4    &   128.6  &  116.3    &  124.4  &   {\pa}4.1   \\
UVES      &  55172.2807   &   0.63 & 116.2    &   115.3  &  115.0    &  115.5  &   {\pa}0.4   \\
UVES      &  55172.3024   &   0.64 & 118.9    &   111.3  & {\pa}99.4 &  109.9  &   {\pa}5.7   \\
UVES      &  55173.2985   &   0.78 & 288.2    &   280.0  &  283.8    &  284.0  &   {\pa}2.3   \\
UVES      &  55173.3201   &   0.78 & 293.6    &   285.3  &  287.7    &  288.9  &   {\pa}2.5   \\
UVES      &  55178.1555   &   0.49 &{\pa}53.9 &{\pa}52.9 & {\pa}49.7 &{\pa}52.2&   {\pa}1.3   \\
UVES      &  55178.1772   &   0.49 &{\pa}52.5 &{\pa}55.5 & {\pa}57.4 &{\pa}55.2&   {\pa}1.4   \\
\hline                            
\rule{0pt}{10pt}LR02             
          &  54748.2657   &   0.11 & 399.2    &   403.6   & 387.6    &  396.8  &  {\pa}4.8 & \rdelim\}{4}{0mm}[134] \\
LR02      &  54748.2873   &   0.12 & 404.6    &   399.6   & 386.3    &  396.8  &  {\pa}5.5   \\
LR02      &  54748.3115   &   0.12 & 393.8    &   395.6   & 377.2    &  388.9  &  {\pa}5.8   \\
LR02      &  54748.3332   &   0.12 & 410.0    &   398.2   & 381.1    &  396.5  &  {\pa}8.4   \\
LR02      &  54749.2121   &   0.25 & 246.2    & $\cdots$  & 228.0    &  237.1  &  {\pa}9.1   \\
LR02      &  54749.2337   &   0.25 & 240.8    &   236.2   & 224.0    &  233.7  &  {\pa}5.0   \\
LR02      &  54837.1223   &   0.01 & 462.8    &   466.0   & 460.3    &  463.0  &  {\pa}1.6 & \rdelim\}{2}{0mm}[157]   \\ 
LR02      &  54837.1440   &   0.01 & 468.2    &   463.3   & 462.9    &  464.8  &  {\pa}1.7   \\
LR02      &  54868.0461   &   0.49 &{\pa}43.0 &{\pa}51.5  &{\pa}28.0 &{\pa}40.9&  {\pa}6.9 & \rdelim\}{2}{0mm}[400]   \\ 
LR02      &  54868.0677   &   0.50 &{\pa}45.8 &{\pa}47.5  &{\pa}28.0 &{\pa}40.4&  {\pa}6.2   \\
LR02      &  55112.3016   &   0.93 & 450.6    &   443.4   & 456.4    &  450.2  &  {\pa}3.8 & \rdelim\}{2}{0mm}[175]   \\ 
LR02      &  55112.3232   &   0.93 & 454.7    &   443.4   & 456.4    &  451.5  &  {\pa}4.1   \\
LR02      &  56210.3545   &   0.25 & 243.6    &   236.4   &  223.7    &  234.6  &  {\pa}5.8   \\
LR02      &  56210.3665   &   0.25 & 243.4    &   236.4   &  214.6    &  231.5  &  {\pa}8.7   \\
LR02      &  56210.3785   &   0.25 & 245.5    &   230.5   &  230.1    &  235.4  &  {\pa}5.1   \\
LR02      &  56217.3299   &   0.26 & 219.8    &   211.2   &  212.1    &  214.4  &  {\pa}2.7   \\
LR02      &  56217.3419   &   0.26 & 221.7    &   207.5   &  198.1    &  209.1  &  {\pa}6.9   \\
LR02      &  56217.3538   &   0.26 & 217.3    &   206.2   &  219.5    &  214.3  &  {\pa}4.1   \\
LR02      &  56243.3377   &   0.03 & 456.1    &   457.2   &  435.2    &  449.5  &  {\pa}7.2 & \rdelim\}{3}{0mm}[140]   \\ 
LR02      &  56243.3497   &   0.03 & 456.0    &   458.8   &  461.5    &  458.8  &  {\pa}1.6   \\
LR02      &  56243.3616   &   0.04 & 455.5    &   457.8   &  438.0    &  450.4  &  {\pa}6.3   \\
LR02      &  56256.2607   &   0.91 & 431.4    &   415.0   &  445.6    &  430.7  &  {\pa}8.8 & \rdelim\}{3}{0mm}[115]   \\ 
LR02      &  56256.2727   &   0.91 & 432.6    &   414.1   &  436.0    &  427.6  &  {\pa}6.8   \\
LR02      &  56256.2846   &   0.91 & 425.5    &   412.7   &  440.4    &  426.2  &  {\pa}8.0   \\
LR02      &  56257.1301   &   0.03 & 460.9    &   463.2   &  457.9    &  460.7  &  {\pa}1.5 & \rdelim\}{3}{0mm}[ 86]   \\ 
LR02      &  56257.1421   &   0.03 & 465.9    &   460.0   &  469.2    &  465.0  &  {\pa}2.7   \\
LR02      &  56257.1541   &   0.04 & 463.1    &   460.8   &  430.1    &  451.3  &      10.6   \\
LR02      &  56277.3081   &   0.96 & 474.3    &   459.5   &  487.0    &  473.6  &  {\pa}7.9 & \rdelim\}{3}{0mm}[146]   \\ 
LR02      &  56277.3201   &   0.96 & 467.9    &   455.6   &  469.3    &  464.3  &  {\pa}4.4   \\
LR02      &  56277.3320   &   0.96 & 475.2    &   459.4   &  493.2    &  475.9  &  {\pa}9.8   \\
LR02      &  56283.0488   &   0.79 & 301.1    &   288.0   &  293.2    &  294.1  &  {\pa}3.8   \\
LR02      &  56283.0608   &   0.80 & 304.0    &   284.3   &  297.5    &  295.3  &  {\pa}5.8   \\
LR02      &  56283.0728   &   0.80 & 305.5    &   287.2   &  302.6    &  298.4  &  {\pa}5.7   \\
LR02      &  56294.1990   &   0.41 &{\pa}90.8 &  {\pa}79.8& {\pa}99.7 &{\pa}90.1&  {\pa}5.8 & \rdelim\}{3}{0mm}[392]   \\ 
LR02      &  56294.2134   &   0.41 &{\pa}87.1 &  {\pa}82.7&  106.6    &{\pa}92.1&  {\pa}7.3   \\
LR02      &  56294.2254   &   0.42 &{\pa}88.1 &  {\pa}81.9& {\pa}94.1 &{\pa}88.0&  {\pa}3.5   \\
LR02      &  56295.1816   &   0.55 &{\pa}63.5 &  {\pa}63.5& {\pa}61.0 &{\pa}62.7&  {\pa}0.8 & \rdelim\}{3}{0mm}[437]   \\ 
LR02      &  56295.1935   &   0.56 &{\pa}67.1 &  {\pa}64.0& {\pa}61.9 &{\pa}64.3&  {\pa}1.5   \\
LR02      &  56295.2055   &   0.56 &{\pa}68.9 &  {\pa}67.9& {\pa}44.8 &{\pa}60.5&  {\pa}7.9   \\
LR02      &  56304.2360   &   0.87 & 378.0    &   366.7   &  381.7    &  375.5  &  {\pa}4.5   \\
LR02      &  56305.2315   &   0.01 & 466.2    &   461.2   &  428.2    &  451.9  &      11.9 & \rdelim\}{3}{0mm}[173]   \\ 
LR02      &  56305.2435   &   0.01 & 473.3    &   464.0   &  468.8    &  468.7  &  {\pa}2.7   \\
LR02      &  56305.2555   &   0.02 & 472.3    &   459.7   & $\cdots$  &  466.0  &  {\pa}3.6   \bigstrut[b]\\
\pagebreak                        
\rule{0pt}{10pt}LR02             
          &  56306.2189   &   0.16 & 367.7    &   366.3   &  359.7    &  364.6  &  {\pa}2.5   \\
LR02      &  56306.2309   &   0.16 & 350.7    &   367.7   &  317.3    &  345.2  &      14.8   \\
LR02      &  56306.2429   &   0.16 & 356.9    &   360.8   &  339.8    &  352.5  &  {\pa}6.4   \\
LR02      &  56308.1546   &   0.44 &{\pa}80.2 &  {\pa}69.0&  119.6    &{\pa}89.6&      15.3 & \rdelim\}{3}{0mm}[323]   \\ 
LR02      &  56308.1666   &   0.44 &{\pa}87.2 &  {\pa}71.3& {\pa}87.7 &{\pa}82.1&  {\pa}5.4   \\
LR02      &  56308.1786   &   0.44 &{\pa}84.0 &  {\pa}66.9& {\pa}71.8 &{\pa}74.2&  {\pa}5.1   \\
LR02      &  56316.2052   &   0.60 &{\pa}94.1 &  {\pa}87.6& {\pa}76.0 &{\pa}85.9&  {\pa}5.3 & \rdelim\}{3}{0mm}[332]   \\ 
LR02      &  56316.2172   &   0.61 &{\pa}92.4 &  {\pa}92.6& {\pa}81.1 &{\pa}88.7&  {\pa}3.8   \\
LR02      &  56316.2292   &   0.61 &{\pa}94.0 &  {\pa}91.6& {\pa}79.0 &{\pa}88.2&  {\pa}4.7   \\
LR02      &  56347.0132   &   0.07 & 444.9    &   441.1   &  420.1    &  435.4  &  {\pa}7.7 & \rdelim\}{3}{0mm}[132]   \\ 
LR02      &  56347.0251   &   0.08 & 441.2    &   438.3   &  462.3    &  447.3  &  {\pa}7.6   \\
LR02      &  56347.0371   &   0.08 & 437.6    &   434.3   &  416.1    &  429.3  &  {\pa}6.7   \\
LR02      &  56349.0214   &   0.37 & 109.2    &  {\pa}88.5&  105.1    &  100.9  &  {\pa}6.3 & \rdelim\}{3}{0mm}[347]   \\ 
LR02      &  56349.0334   &   0.37 & 107.5    &  {\pa}92.8&  108.1    &  102.8  &  {\pa}5.0   \\
LR02      &  56349.0453   &   0.37 &{\pa}99.8 &  {\pa}91.9& {\pa}95.1 &{\pa}95.6&  {\pa}2.3   \\
LR02      &  56352.0241   &   0.80 & 289.7    &   271.1   &  279.1    &  280.0  &  {\pa}5.4   \\
LR02      &  56352.0360   &   0.80 & 291.9    &   269.7   &  282.6    &  281.4  &  {\pa}6.4   \\
LR02      &  56352.0480   &   0.80 & 296.8    &   278.4   &  280.4    &  285.2  &  {\pa}5.8   \\
LR02      &  56356.0044   &   0.38 &{\pa}90.7 &  {\pa}83.9& {\pa}91.5 &{\pa}88.7&  {\pa}2.4 & \rdelim\}{3}{0mm}[339]   \\ 
LR02      &  56356.0163   &   0.38 &{\pa}95.3 &  {\pa}81.2& {\pa}90.8 &{\pa}89.1&  {\pa}4.2   \\
LR02      &  56356.0283   &   0.38 &{\pa}92.6 &  {\pa}79.0& {\pa}65.4 &{\pa}79.0&  {\pa}7.9   \bigstrut[b]\\
\hline                           
\rule{0pt}{8pt}& &              
& \ion{Si}{iii} & \ion{He}{i} & \ion{He}{i} & \multicolumn{2}{c}{     } &
\\                               
 &  &                            
& $\lambda$4552 & $\lambda$4713 & $\lambda$4922               \\
\hline                           
\rule{0pt}{8pt}LR03      
          &  54755.1987   & 0.12 & 387.0 & 399.3 & 402.7 & 396.3 &   4.8 \\
LR03      &  54810.2265   & 0.10 & 412.4 & 423.2 & 433.8 & 423.1 &   6.2 \\
LR03      &  54810.2481   & 0.11 & 410.5 & 423.7 & 425.3 & 419.8 &   4.7 \\
LR03      &  54810.2699   & 0.11 & 409.6 & 422.1 & 419.7 & 417.1 &   3.8 \\
LR03      &  54810.2915   & 0.11 & 404.9 & 415.6 & 419.6 & 413.4 &   4.4 \\
LR03      &  54810.3248   & 0.12 & 395.2 & 414.2 & 412.6 & 407.3 &   6.1 \\
LR03      &  54810.3465   & 0.12 & 394.2 & 427.0 & 413.5 & 411.6 &   9.5 \\
\hline
\end{longtable}
\end{center}

\twocolumn

\footnotesize
\onecolumn
\begin{center}
\begin{longtable}{lccccccclr}
  \caption{\label{t_full_obs2} Log of spectroscopic observations of VFTS~652;  details are
    as for Table~\ref{t_full_obs1}.\hfill}\\
\hline
\hline
\multicolumn{1}{c}{\rule{0pt}{8pt}Setting} & MJD \T& Orbital &
\multicolumn{5}{c}{Primary radial velocities (\kms)} &
\multicolumn{2}{c}{Secondary} \\
 & & Phase
& \ion{He}{i} & \ion{He}{i} & \ion{Si}{iii} & \multicolumn{2}{c}{     } &
\multicolumn{2}{c}{RV (\kms)} \\
 & &
& $\lambda$4387 & $\lambda$4471 & $\lambda$4552 & Mean & s.e. &
\multicolumn{2}{c}{\heii\ $\lambda$4541} \\
\hline
\endfirsthead
\caption[]{\textit{continued}}\\
\hline
\hline
\multicolumn{1}{c}{\rule{0pt}{8pt}Setting} & MJD \T& Orbital &
\multicolumn{5}{c}{Primary radial velocities (\kms)} &
\multicolumn{2}{c}{Secondary} \\
 & & Phase
& \ion{He}{i} & \ion{He}{i} & \ion{Si}{iii} & \multicolumn{2}{c}{     } &
\multicolumn{2}{c}{RV (\kms)} \\
 & &
& $\lambda$4387 & $\lambda$4471 & $\lambda$4552 & Mean & s.e. &
\multicolumn{2}{c}{\heii\ $\lambda$4541} \\
\hline
\endhead
\hline
\endfoot
\hline
\hline
\endlastfoot
\rule{0pt}{10pt}UVES      
          & 54791.3055  &   0.69 &     179.8  &     171.1 &       160.4  &     170.4 & {\pa}5.6   \\
UVES      & 54791.3271  &   0.69 &     173.0  &     172.4 &       179.9  &     175.1 & {\pa}2.4   \\
UVES      & 54792.1644  &   0.79 &     293.6  &     299.9 &       316.2  &     303.2 & {\pa}6.7   \\
UVES      & 54792.1860  &   0.79 &     301.7  &     293.3 &       320.1  &     305.0 & {\pa}7.9   \\
UVES      & 54847.1501  &   0.19 &     335.5  &     321.2 &       331.8  &     329.5 & {\pa}4.3   \\
UVES      & 54847.1717  &   0.19 &     322.0  &     317.2 &       320.1  &     319.8 & {\pa}1.4   \\
UVES      & 54892.0909  &   0.42 & {\pa}93.2  & {\pa}91.4 &   {\pa}67.4  & {\pa}84.0 & {\pa}8.3   \\
UVES      & 54892.1125  &   0.43 &     113.5  & {\pa}91.4 &   {\pa}82.6  & {\pa}95.8 & {\pa}9.2   \\
UVES      & 54894.0276  &   0.65 &     139.2  &     135.2 &       142.3  &     138.9 & {\pa}2.0   \\
UVES      & 54894.0492  &   0.65 &     137.8  &     132.6 &       125.4  &     131.9 & {\pa}3.6   \\
UVES      & 54896.0284  &   0.88 &     399.2  &     398.2 &       398.0  &     398.5 & {\pa}0.4   \\
UVES      & 54896.0504  &   0.88 &     397.8  &     396.9 &       399.3  &     398.0 & {\pa}0.7   \\
UVES      & 54897.0255  &   1.00 &     457.4  &     480.1 &    $\cdots$  &     468.7 &     11.3   \\
UVES      & 54898.0463  &   0.12 &     408.6  &     408.9 &       400.6  &     406.0 & {\pa}2.7   \\
UVES      & 54898.0687  &   0.12 &     401.9  &     404.9 &       377.1  &     394.6 & {\pa}8.8   \\
UVES      & 55201.1762  &   0.41 & {\pa}91.8  & {\pa}87.4 &   {\pa}78.7  & {\pa}85.9 & {\pa}3.9   \\
UVES      & 55201.1984  &   0.41 & {\pa}89.1  & {\pa}88.7 &   {\pa}94.2  & {\pa}90.7 & {\pa}1.8   \\
\hline                          
\rule{0pt}{10pt}LR02            
          & 54804.0935  &   0.18 &     343.7  &     342.4 &     346.1  &    344.1 & {\pa}1.1  & \rdelim\}{6}{8mm}[232] \\
LR02      & 54804.1151  &   0.18 &     338.2  &     331.8 &     347.4  &    339.1 & {\pa}4.5  \\
LR02      & 54804.1368  &   0.18 &     338.2  &     339.8 &     344.8  &    340.9 & {\pa}2.0  \\
LR02      & 54804.1584  &   0.19 &     336.9  &     335.8 &     338.3  &    337.0 & {\pa}0.7  \\
LR02      & 54804.1801  &   0.19 &     332.8  &     333.1 &     334.4  &    333.4 & {\pa}0.5  \\
LR02      & 54804.2016  &   0.19 &     324.7  &     334.5 &     329.2  &    329.4 & {\pa}2.8  \\
LR02      & 54836.2280  &   0.92 &     431.7  &     431.4 &     440.8  &    434.6 & {\pa}3.1  & \rdelim\}{4}{8mm}[212] \\
LR02      & 54836.2497  &   0.92 &     429.0  &     430.1 &     438.2  &    432.4 & {\pa}2.9  \\
LR02      & 54836.2758  &   0.92 &     430.3  &     430.1 &     439.5  &    433.3 & {\pa}3.1  \\
LR02      & 54836.2974  &   0.93 &     442.5  &     436.8 &     436.9  &    438.7 & {\pa}1.9  \\
LR02      & 54867.0978  &   0.51 & {\pa}60.6  & {\pa}59.5 & {\pa}52.7  &{\pa}57.6 & {\pa}2.5  & \rdelim\}{2}{8mm}[329] \\
LR02      & 54867.1195  &   0.52 & {\pa}63.4  & {\pa}63.5 & {\pa}61.8  &{\pa}62.9 & {\pa}0.5  \\
LR02      & 55108.3179  &   0.60 & {\pa}83.7  & {\pa}86.1 & {\pa}82.6  &{\pa}84.1 & {\pa}1.0  & $\phantom{\}\,}\cdots$ \\
LR02      & 55114.3094  &   0.29 &     183.9  &     180.4 &     186.4  &    183.6 & {\pa}1.7  & \rdelim\}{2}{8mm}[258] \\
LR02      & 55114.3310  &   0.30 &     183.9  &     179.0 &     191.6  &    184.8 & {\pa}3.7  \\
LR02      & 56210.3545  &   0.90 &     419.5  &     410.1 &     418.9  &    416.2 & {\pa}3.0 & \rdelim\}{3}{8mm}[247] \\
LR02      & 56210.3665  &   0.90 &     417.2  &     404.2 &     424.1  &    415.2 & {\pa}5.8 \\
LR02      & 56210.3785  &   0.90 &     410.7  &     411.4 &     426.0  &    416.0 & {\pa}5.0 \\
LR02      & 56217.3299  &   0.71 &     192.7  &     182.5 &     195.4  &    190.2 & {\pa}3.9 & \rdelim\}{3}{8mm}[338] \\
LR02      & 56217.3419  &   0.71 &     196.0  &     181.8 &     199.9  &    192.6 & {\pa}5.5 \\
LR02      & 56217.3538  &   0.71 &     192.2  &     185.4 &     203.8  &    193.8 & {\pa}5.4 \\
LR02      & 56243.3377  &   0.74 &     229.1  &     212.2 &     236.8  &    226.0 & {\pa}7.3 & \rdelim\}{3}{8mm}[278] \\
LR02      & 56243.3497  &   0.74 &     232.2  &     215.3 &     237.0  &    228.2 & {\pa}6.6 \\
LR02      & 56243.3616  &   0.74 &     233.7  &     217.6 &     237.8  &    229.7 & {\pa}6.2 \\
LR02      & 56256.2607  &   0.24 &     267.8  &     240.1 &     266.4  &    258.1 & {\pa}9.0 & \rdelim\}{3}{8mm}[297] \\
LR02      & 56256.2727  &   0.24 &     260.6  &     236.7 &     269.7  &    255.7 & {\pa}9.8 \\
LR02      & 56256.2846  &   0.24 &     256.4  &     242.9 &     264.0  &    254.4 & {\pa}6.2 \\
LR02      & 56257.1301  &   0.34 &     151.9  &     128.2 &     145.2  &    141.8 & {\pa}7.1 & \rdelim\}{3}{8mm}[297] \\
LR02      & 56257.1421  &   0.34 &     142.8  &     131.9 &     139.3  &    138.0 & {\pa}3.2 \\
LR02      & 56257.1541  &   0.34 &     138.4  &     128.7 &     142.2  &    136.4 & {\pa}4.0 \\
LR02      & 56277.3081  &   0.69 &     174.9  &     173.2 &     187.8  &    178.6 & {\pa}4.6 & \rdelim\}{3}{8mm}[376] \\
LR02      & 56277.3201  &   0.69 &     183.0  &     172.7 &     192.4  &    182.7 & {\pa}5.7 \\
LR02      & 56277.3320  &   0.69 &     187.1  &     177.6 &     180.4  &    181.7 & {\pa}2.8 \\
LR02      & 56283.0488  &   0.36 &     137.7  &     133.8 &     144.1  &    138.5 & {\pa}3.0 & \rdelim\}{3}{8mm}[348] \\
LR02      & 56283.0608  &   0.36 &     137.3  &     126.7 &     129.7  &    131.2 & {\pa}3.2 \\
LR02      & 56283.0728  &   0.36 &     135.2  &     133.3 &     132.2  &    133.6 & {\pa}0.9 \\
LR02      & 56294.1990  &   0.66 &     145.7  &     149.0 &     152.5  &    149.1 & {\pa}2.0 & \rdelim\}{3}{8mm}[294] \\
LR02      & 56294.2134  &   0.66 &     149.4  &     148.5 &     150.0  &    149.3 & {\pa}0.4 \\
LR02      & 56294.2254  &   0.66 &     153.9  &     149.8 &     147.3  &    150.3 & {\pa}1.9 \\
LR02      & 56295.1816  &   0.77 &     278.2  &     261.0 &     291.5  &    276.9 & {\pa}8.8 & \rdelim\}{3}{8mm}[245] \\
LR02      & 56295.1935  &   0.77 &     277.7  &     264.3 &     283.6  &    275.2 & {\pa}5.7 \\
LR02      & 56295.2055  &   0.77 &     283.2  &     264.8 &     294.7  &    280.9 & {\pa}8.7 \\
LR02      & 56304.2360  &   0.83 &     342.8  &     332.2 &     357.8  &    344.3 & {\pa}7.4 & $\phantom{\}\,\,}253$ \\
LR02      & 56305.2315  &   0.94 &     444.5  &     438.9 &     436.3  &    439.9 & {\pa}2.4 & \rdelim\}{3}{8mm}[156] \\
LR02      & 56305.2435  &   0.94 &     450.1  &     442.5 &     449.7  &    447.4 & {\pa}2.5 \\
LR02      & 56305.2555  &   0.94 &     449.1  &     442.7 &     458.6  &    450.1 & {\pa}4.6 \bigstrut[b]\\
\pagebreak                      
LR02      & 56306.2189  &   0.06 &     449.4  &     444.9 &     457.9  &    450.7 & {\pa}3.8 & \rdelim\}{3}{8mm}[215] \\
LR02      & 56306.2309  &   0.06 &     448.5  &     449.5 &     454.0  &    450.7 & {\pa}1.7 \\
LR02      & 56306.2429  &   0.06 &     451.8  &     446.3 &     453.1  &    450.4 & {\pa}2.1 \\
LR02      & 56308.1546  &   0.28 &     196.2  &     175.1 &     201.3  &    190.9 & {\pa}8.0 & \rdelim\}{3}{8mm}[333] \\
LR02      & 56308.1666  &   0.28 &     196.8  &     178.2 &     190.6  &    188.5 & {\pa}5.5 \\
LR02      & 56308.1786  &   0.28 &     198.7  &     178.4 &     204.7  &    193.9 & {\pa}8.0 \\
LR02      & 56316.2052  &   0.22 &     300.9  &     267.0 &     293.7  &    287.2 &     10.3 & \rdelim\}{3}{8mm}[236] \\
LR02      & 56316.2172  &   0.22 &     292.0  &     268.7 &     290.9  &    283.9 & {\pa}7.6 \\
LR02      & 56316.2292  &   0.22 &     292.9  &     269.1 &     282.8  &    281.6 & {\pa}6.9 \\
LR02      & 56347.0132  &   0.81 &     319.8  &     304.4 &     327.5  &    317.2 & {\pa}6.8 & \rdelim\}{3}{8mm}[255] \\
LR02      & 56347.0251  &   0.81 &     316.2  &     311.0 &     325.4  &    317.5 & {\pa}4.2 \\
LR02      & 56347.0371  &   0.81 &     321.1  &     311.9 &     326.4  &    319.8 & {\pa}4.2 \\
LR02      & 56349.0214  &   0.04 &     449.7  &     447.3 &     463.5  &    453.5 & {\pa}5.0 & \rdelim\}{3}{8mm}[201] \\
LR02      & 56349.0334  &   0.04 &     450.8  &     442.9 &     459.8  &    451.2 & {\pa}4.9 \\
LR02      & 56349.0453  &   0.04 &     457.4  &     448.2 &     454.0  &    453.2 & {\pa}2.7 \\
LR02      & 56352.0241  &   0.39 &     105.5  & {\pa}93.6 &     111.1  &    103.4 & {\pa}5.2 & \rdelim\}{3}{8mm}[367] \\
LR02      & 56352.0360  &   0.39 &     103.0  & {\pa}99.4 &     107.0  &    103.1 & {\pa}2.2 \\
LR02      & 56352.0480  &   0.39 &     103.2  & {\pa}99.7 & {\pa}96.9  &{\pa}99.9 & {\pa}1.8 \\
LR02      & 56356.0044  &   0.85 &     370.5  &     368.2 &     379.4  &    372.7 & {\pa}3.4 & \rdelim\}{3}{8mm}[274] \\
LR02      & 56356.0163  &   0.85 &     380.6  &     367.4 &     386.5  &    378.2 & {\pa}5.6 \\
LR02      & 56356.0283  &   0.86 &     379.5  &     368.2 &     384.5  &    377.4 & {\pa}4.8 \bigstrut[b]\\
\hline
\rule{0pt}{8pt}& &
& \ion{Si}{iii} & \ion{He}{i} & \ion{He}{i} & \multicolumn{2}{c}{     } &
\\
 &  &
& $\lambda$4552 & $\lambda$4713 & $\lambda$4922 \\
\hline
\rule{0pt}{8pt}LR03      & 54808.1322 & 0.65 & 135.0 & 136.2 & 136.3 & 135.8 & {\pa}0.4   & \rdelim\}{6}{8mm}[340] \\
               LR03      & 54808.1538 & 0.65 & 138.1 & 140.6 & 140.2 & 139.6 & {\pa}0.8   \\
               LR03      & 54808.1755 & 0.65 & 142.6 & 143.2 & 142.4 & 142.7 & {\pa}0.2   \\
               LR03      & 54808.1971 & 0.66 & 142.6 & 149.8 & 144.3 & 145.6 & {\pa}2.2   \\
               LR03      & 54808.2189 & 0.66 & 145.9 & 146.6 & 149.4 & 147.3 & {\pa}1.1   \\
               LR03      & 54808.2405 & 0.66 & 149.0 & 150.9 & 153.3 & 151.1 & {\pa}1.2   \\
\hline
\end{longtable}
\end{center}

\twocolumn

\end{document}